\documentclass[aps,pra,twocolumn,floatfix,superscriptaddress,longbibliography]{revtex4-1}

\usepackage{xcolor}
\usepackage{appendix}
\usepackage{graphicx,times}
\usepackage{amstext}
\usepackage{amsmath}            
\usepackage{amssymb}            
\usepackage{graphicx}           
\usepackage{latexsym}
\usepackage{bm}

\usepackage{etoolbox}
\usepackage{breqn}

\usepackage[FIGTOPCAP,raggedright,nooneline]{subfigure}

\makeatletter
\let\cat@comma@active\@empty
\makeatother

\usepackage{url}
\usepackage[colorlinks]{hyperref}
\hypersetup{%
    plainpages=true,
    breaklinks=true,
    hypertexnames=false,
    pageanchor=true,
    colorlinks=true,
    linkcolor={blue},
    citecolor={red},
    urlcolor={blue},
    anchorcolor={black}
}

\newcommand{\beq}{\begin{eqnarray}}
\newcommand{\eeq}{\end{eqnarray}}

\newcommand{\nn}{\nonumber}

\DeclareMathOperator{\tr}{tr}

\newtheorem{theorem}{Theorem}
\newtheorem{lemma}{Lemma}

\def\<{\langle}
\def\>{\rangle}

\def\II{\mathcal{I}}

\def\<{\langle}
\def\>{\rangle}

\def \info#1{}

\def \info#1{}

\newcommand{\Smax}{S_{\rm max}}
\newcommand{\Smin}{S_{\rm min}}

\begin{document}

\title{Einstein-Podolsky-Rosen steering: Its geometric quantification and
witness}

\author{Huan-Yu Ku}
\affiliation{Department of Physics, National Cheng Kung
University, Tainan 701, Taiwan}
\author{Shin-Liang Chen}
\affiliation{Department of Physics, National Cheng Kung
University, Tainan 701, Taiwan}
\author{Costantino Budroni}
\affiliation{Institute for Quantum Optics and Quantum Information
(IQOQI), Boltzmanngasse 3 1090 Vienna, Austria}
\author{Adam Miranowicz}
\affiliation{CEMS, RIKEN, 351-0198 Wako-shi, Japan}
\affiliation{Faculty of Physics, Adam Mickiewicz University,
61-614 Pozna\'n, Poland}
\author{Yueh-Nan Chen}
\email{yuehnan@mail.ncku.edu.tw} \affiliation{Department of
Physics, National Cheng Kung University, Tainan 701, Taiwan}
\affiliation{CEMS, RIKEN, 351-0198 Wako-shi, Japan}
\affiliation{Physics Division, National Center for Theoretical
Sciences, Hsinchu 300, Taiwan}
\author{Franco Nori}
\affiliation{CEMS, RIKEN, 351-0198 Wako-shi, Japan}
\affiliation{Department of Physics, The University of Michigan,
Ann Arbor, Michigan 48109-1040, USA}
\date{\today}

\begin{abstract}
We propose a measure of quantum steerability, namely a convex
steering monotone, based on the trace distance between a given
assemblage and its corresponding closest assemblage admitting a
local-hidden-state (LHS) model. We provide methods to estimate
such a quantity, via lower and upper bounds, based on semidefinite
programming. One of these upper bounds has a clear geometrical
interpretation as a linear function of rescaled Euclidean
distances in the Bloch sphere between the normalized quantum
states of: (i) a given assemblage and (ii) an LHS assemblage. For
a qubit-qubit quantum state, these ideas also allow us to
visualize various steerability properties of the state in the
Bloch sphere via the so-called LHS surface. In particular, some
steerability properties can be obtained by comparing such an LHS
surface with a corresponding quantum steering ellipsoid. Thus, we
propose a witness of steerability corresponding to the difference
of the volumes enclosed by these two surfaces. This witness (which
reveals the steerability of a quantum state) enables one to find an
optimal measurement basis, which can then be used to determine the
proposed steering monotone (which describes the steerability of an
assemblage) optimized over all mutually-unbiased bases.
\end{abstract}

\maketitle


\section{Introduction}

Quantum entanglement~\cite{Einstein35}, Einstein-Podolsky-Rosen
(EPR) steering~\cite{Schrodinger35}, and Bell
nonlocality~\cite{Bell64} are different forms of quantum
nonlocality~\cite{Wiseman07}. These quantum correlations are
powerful resources for quantum engineering, quantum cryptography,
quantum communication, and quantum information
processing~\cite{Horodecki09RMP,Brunner14RMP,Gallego15}. Taking an
operational perspective~\cite{Wiseman07}, EPR steering can certify
the entanglement between two systems when one of the measurements
is untrusted, i.e., no assumptions are made on the functioning of
the measurement device. On the other hand, Bell nonlocality can
certify the entanglement with the untrusted measurements on both
sides. One can also certify an entangled state by performing
quantum state tomography with all-trusted measurement devices.
Thus, EPR steering is a form of quantum correlation, which can be
classified between entanglement and Bell nonlocality, in the
following meaning: it certifies the entanglement between two
systems assuming trusted measurements only on one of
these~\cite{Wiseman07}. Eighty years of research on EPR steering
has resulted in many experimental demonstrations~\cite{Ou92,
Hald99,Bowen03, Howell04,Saunders10, Wittmann12, Bennet12,
Handchen12, Smith12,Steinlechner13, Schneeloch13,Su13,
Sun16, Cavalcanti09} and various applications~\cite{Reid89,
Pusey13,Walborn11,Kogias15,Costa16}, which include multipartite
quantum steering~\cite{He13, Li_Che_Ming15, Xiang17,Milne14,
Cheng16}, the correspondence with measurement
incompatibility~\cite{Cavalcanti16,
Uola14,Quintino14,Shin-Liang16c,Uola15}, one-way
steering~\cite{Sun16, Wollmann16,Skrzypczyk14,SDPreview17}, one
sided device-independent processing in quantum key
distribution~\cite{Branciard12}, continuous-variable EPR
steering~\cite{Tatham12,Qiongyi15,Wang15,Xiang17}, as well as
temporal~\cite{Yueh-Nan14, Shin-Liang16,Bartkiewicz16a,
Bartkiewicz16b,Ku16,Che-Ming15} and spatio-temporal
steering~\cite{Chen2017}.

In recent years, several measures of steering, such as steerable
weight~\cite{Skrzypczyk14}, steering robustness~\cite{Piani15},
steering fraction~\cite{Hsieh16}, steering cost~\cite{Das17},
intrinsic steerability \cite{Eneet17a}, as well as the relative
entropy of steering~\cite{Gallego15,Eneet17b} have been proposed
(see also the review~\cite{SDPreview17}). All these quantifiers
are monotones under one-way local operations assisted by classical
communication (one-way LOCCs)~\cite{Gallego15}. More recently,
several works using the geometrical approaches to steering have
been considered, such as depicting quantum correlations for
two-qubit states~\cite{Jevtic14,Jevtic15} and a geometrical approach
to witness steering~\cite{Nguyen16a,Nguyen16b}. Here, we would
like to use the \emph{consistent steering robustness} (CSR)
introduced by Cavalcanti~\emph{et al.}~\cite{Cavalcanti16} and the
\emph{quantum steering ellipsoid} (QSE) introduced by Jevtic
\emph{et al}.~\cite{Jevtic14,Jevtic15} to construct such a
geometric witness. The QSE provides a visualization and geometric
representation of any two-qubit
state~\cite{McCloskey16,Cheng16,Bowles16,Milne14}. Specifically,
the QSE for a given two-qubit state corresponds to the set of all
Bloch vectors of one qubit (say Bob), which can be prepared by
another qubit (say Alice) by considering all possible projective
measurements on her qubit~\cite{Jevtic14,Jevtic15}.

In this work, we propose a distance between assemblages based on
the trace distance between single elements. Given an assemblage, a
trace-distance measure of steerability is then proposed as the
distance to the closest unsteerable assemblage. Here, we prove
that the consistent trace-distance measure of steerability is a
\emph{convex steering monotone}, with respect to restricted
one-way LOCCs introduced in Ref.~\cite{Eneet17a}. We note that our
proposal is reminiscent of other distance-based measures of
various quantum phenomena. These include  ``nonclassical
distance'' for quantifying the quantumness of optical
fields~\cite{Hillery87,Adam15}, distance-based measures of
entanglement~\cite{Vedral97}, trace-distance measures of
coherence~\cite{Rana16}, or trace-distance measures quantifying
Bell nonlocality~\cite{Brito17}.

In order to estimate the proposed steering monotone, we provide
lower and upper bounds that can be efficiently computed by
semidefinite programs (SDPs)~\cite{SDP}. Specifically, a lower
bound is obtained via an operator-norm distance, whereas a few
upper bounds are found by applying various known steering
measures~\cite{Cavalcanti16,Skrzypczyk14,Piani15,Bavaresco17}.

Moreover, we introduce the local-hidden-state (LHS) surface as a
way of visualizing steerability properties of a two-qubit quantum
state in the Bloch sphere. In particular, these notions connect to
the QSE and provide a witness of steerability based on the
different volumes enclosed by the two surfaces. This steerability
witness enables finding an optimal measurement
basis~\cite{McCloskey16}. Thus, this is particularly important for
calculating the proposed steering monotone optimized over all
mutually-unbiased bases. To illustrate the usefulness of LHS
surfaces, we provide the explicit solution of the LHS surface for
the Werner states. Moreover, we present a few upper and lower
bounds of the steerability measure for the Werner, Horodecki, and
rank-2 Bell-diagonal states~\cite{Werner89,Adam08}. Note that this
approach, despite some resemblance, essentially differs from,
e.g., the relative entropy of
entanglement~\cite{Vedral97,Vedral98,Adam08} and the nonclassical
distance~\cite{Hillery87,Adam15} used for quantifying the
quantumness of bosonic systems.

This paper is organized as follows. In Sec.\ref{sec:prel} we
summarize the basic notions concerning EPR steering. In
Sec.~\ref{sec:quanti} we introduce a steering quantifier; we also
prove that it is a monotone under restricted one-way LOCCs, and
provide computable lower and upper bounds for it. In
Sec.~\ref{sec:witness} we introduce a steering witness based on
the notion of LHS surface and discuss its properties. In
Sec.~\ref{sec:appli} we apply our results to several interesting
examples. Finally, in Sec.~\ref{sec:conclu}, we provide the
conclusions and outlook for our work.

\section{Preliminary notions}\label{sec:prel}

EPR steering can be operationally defined as the success of the
following task~\cite{Wiseman07}: One party, say Alice, tries to
convince another party, say Bob, that they share an entangled
state $\rho_{AB}$. To accomplish this task, Bob asks Alice to
perform some measurements, described by positive-operator valued
measures (POVMs) $A_{a|x}$ with $A_{a|x}\geq 0$, satisfying
$\sum_{a}A_{a|x}=\openone$, where $x$ denotes the basis of the
measurement, $a$ is its outcome, and $\openone$ is an unit
operator. Bob's measurements are assumed to be fully characterized
by quantum mechanics. Therefore, he can perform quantum state
tomography and obtain the unnormalized quantum states
$\sigma_{a|x}=\text{tr}_{A}(\rho _{AB}A_{a|x}\otimes \openone)$.
In particular, any $\{A_{a|x}\}_{a,x}$ gives rise to a collection
of unnormalized quantum states $\{\sigma _{a|x}\}_{a,x}$, which
are termed as an \textit{assemblage}. An assemblage also includes
the information of Alice's marginal statistics
$p(a|x)=\text{tr}(\sigma _{a|x})$.

The assemblage $\{\sigma _{a|x}\}_{a,x}$ is unsteerable if it admits an LHS
model:
\begin{equation}
\sigma _{a|x}=\sigma _{a|x}^{\text{US}}=\sum_{\lambda }p(\lambda
)p(a|x,\lambda )\sigma _{\lambda }~\quad\forall a,x.  \label{lhs}
\end{equation}
An LHS model can be understood as follows: Alice sends a
preexisting quantum state $\sigma _{\lambda }$ according to her
input $x$ and outcome $a$ with a probability distribution
$p(\lambda )$ and a conditional probability distribution
$p(a|x,\lambda )$. In this sense, the assemblage, received by Bob, is
just a classical postprocessing of the set of states $\{\sigma
_{\lambda }\}_{\lambda}$, which is clearly independent of Alice's
measurements. Likewise, a quantum state $\rho _{AB}$ is called
\emph{steerable} if the given assemblage does not admit an LHS
model. Such a state is necessarily entangled, but the converse is
not true~\cite{Wiseman07}.

In the context of a resource theory of steering~\cite{Gallego15}, the most general free operation for EPR steering is a stochastic one-way LOCC, defined as follows. Given an assemblage $\{\sigma_{a|x}\}_{a,x}$, Bob performs a quantum measurement on his system. The measurement is described by a completely positive trace-nonincreasing map $\varepsilon$ defined by
\begin{equation}
\begin{split}
\varepsilon (\sigma_{B}) :=\sum_\omega K_\omega (\sigma_{B})K^\dagger_\omega,
~~~\text{such~~~~that}~~\sum_\omega K^\dagger_\omega K_\omega \leq \openone,
\end{split}
\end{equation}
for the reduced state $\sigma_{B}$ of Bob,
where $K_\omega$ is the Kraus operator associated with a classical outcome $\omega$. In the most general case, the set of classical outcomes is a coarse graining of the set of possible $\omega$ (quantum instruments may be defined by more than one Kraus operator), but, as we discuss below, there is no loss of generality by considering that each outcome $\omega$ is associated with a single Kraus operator $K_\omega$.

After such an operation, Bob communicates with Alice obtaining a classical result $\omega$ prior to her measurement. She applies a local deterministic wiring map $W_{\omega}$, defined explicitly below, described by the normalized conditional probability distributions: $p(x|x',\omega)$, describing the generation of any initial input $x$ from final input $x'$ and Bob's result $\omega$, and $p(a'|a,x,x',\omega)$, describing the generation of Alice's final outcome $a'$ from $a$, $x$, $x'$, and $\omega$.
The final assemblage with input $x'$ and outcomes $\omega, a'$ becomes
\begin{equation}
\{\sigma_{a'|x'}^\omega\}_{a',x'}:=M_\omega(\{\sigma_{a|x}\}_{a,x}).
\end{equation}
Here, $M_{\omega}(\{\sigma_{a|x}\}_{a,x}):=K_\omega W_{\omega}(\{\sigma_{a|x}\}_{a,x}) K_\omega^\dagger$ is a subchannel of the map $M:=\sum_\omega M_\omega$ when Bob post-selects the $\omega$th outcome with probability $p(\omega)$=Tr[$M_\omega (\sum_a \{\sigma_{a|x}\}_{a,x})$], while
\begin{equation}
W_\omega(\{\sigma_{a|x}\}_{a,x})=\{\sum_{a,x}p(x|x',\omega)p(a'|a,x,x',\omega)\sigma_{a|x}\}_{a',x'}
\end{equation}
is a deterministic wiring map.
We recall that a function $S$ is a
steering monotone (see Ref.~\cite{Gallego15}) if it is zero for
unsteerable assemblages and it is a monotone, i.e., it does not
increase (on average), under one-way LOCCs, i.e.,
\begin{equation}
\sum_\omega p(\omega)S \left(\frac{M_\omega (\{\sigma_{a|x}\}_{a,x})}{p(\omega)} \right)\leq S(\{\sigma_{a|x}\}_{a,x}),
\end{equation}
for a given assemblage $\{\sigma_{a|x}\}_{a,x}$. Note that the particular case when $\sum_{\omega}p(\omega)=1$ or $\sum_{\omega}K^\dagger_\omega K_\omega=\openone$, is called a \rm{deterministic} one-way LOCC. Otherwise, this is a \rm{stochastic} one-way LOCC.
Finally, we note that the use of a coarse-grained set of classical outcomes, simply implies the equality of some of Alice wirings, i.e., $W_\omega= W_\omega'$, if $\omega$ and $\omega'$ are coarse-grained into the same classical outcome.

In the following, we consider a \emph{restricted} set of one-way
LOCC, which has been proposed in Ref.~\cite{Eneet17a}. This
restriction consists of requiring that Alice's choice of wiring
does not depend on the classical outcome $\omega$ obtained by Bob
via local operations. This restriction can be motivated by
practical reasons~\cite{Eneet17a}: Given a spatial separation
between Alice and Bob, the protocol may be more efficient if Alice
directly applies her operation on her system instead of waiting
for communication with Bob. This ``restriction hypothesis''
translates into the condition $p(x|x',\omega)=p(x|x')$, and hence
into the condition
$$W_\omega(\{\sigma_{a|x}\}_{a,x})=\{\sum_{a,x}p(x|x')p(a'|a,x,x',\omega)\sigma_{a|x}\}_{a',x'}.$$

\section{Geometric quantifiers of steerability}\label{sec:quanti}

\subsection{Trace-distance steerability  measure}

The (quantum) trace distance is a metric to distinguish two
density operators $\rho $ and $\rho^{\prime}$, i.e., $D_{Q}(\rho
,\rho ^{\prime }):=\frac{1}{2}\|\rho -\rho^{\prime }\|$, where $\|
X \|:= \tr[|X|]$ is the trace norm. When $\rho$ and $\rho^{\prime
}$ commute, the trace distance reduces to the classical trace
distance, i.e., the Kolmogorov distance~\cite{Fuchs99}, which can
be defined as $D_{C}(P,P^{\prime
}):=\frac{1}{2}\sum_{x}|P(x)-P^{\prime }(x)|$ for two probability
distributions $P$ and $P^{\prime}$.

One can easily prove the properties of a metric, i.e., it is (i)
non-negative, (ii) symmetric, (iii) vanishes if and only if
$\rho=\rho^{\prime}$, and (iv) satisfies the triangle inequality.
Similarly, we can define the distance between two assemblages as
\begin{equation}
D_{A}(\{\sigma _{a|x}\}_{a,x},\{\sigma_{a|x}^{\prime }\}_{a,x})
= \sum_{a,x}p(x)D(\sigma_{a|x},\sigma_{a|x}^{\prime
}), \label{eq:dist_assm}
\end{equation}
where $D(a,b):=\frac{1}{2}\|a-b\|$, and $\{\sigma _{a|x}\}_{a,x}$
and $\{\sigma_{a|x}^{\prime }\}_{a,x}$ are two assemblages with
the same number of inputs $N_x$. In general, $p(x)$ can be chosen
to be uniform with respect to the number of measurement settings,
i.e., $\frac{1}{N_{x}}$. As for $D_Q$, one can easily prove that
$D_A$ satisfies all the properties of a metric (cf.
Appendix~\ref{app:metric}). Note that trace-distance between two
assemblages is first introduced by Kaur~\emph{et
al.}~\cite{Eneet17b}. Nevertheless, our definition is different
from theirs.

In the following, we want to introduce a measure of steerability
based on the distance of a given assemblage from the set of
unsteerable states. Several convex steering monotones have been
introduced, with different properties and different
interpretations. Our goal here is to introduce a different measure based
on the trace-distance between assemblages. We introduce a
quantifier, called {\it consistent trace-distance measure of
steerability}, defined as the minimal trace distance to the
``consistent'' unsteerable assemblage \cite{Cavalcanti16}, namely,
\begin{equation}
\begin{split}
S_{\rm TD}(\{\sigma_{a|x}\}_{a,x} ):=\min \Big\lbrace D_{A}({\{\sigma_{a|x}\}_{a,x},\{\rho_{a|x}\}_{a,x}}) \Big| \\
{\{\rho_{a|x}\}_{a,x}\in \text{LHS}}, \  \sum_a \rho_{a|x} = \sum_a \sigma_{a|x}, \ \forall x\ \Big\rbrace.
\end{split}
\end{equation}
where ${\rm LHS}$ denotes the set of unsteerable assemblages,
i.e., those admitting an LHS model given by Eq.~\eqref{lhs}. In
Appendix~\ref{app:mono}, we prove that $S_{\rm TD}$ is a
restricted \emph{convex steering monotone}.

Unfortunately, it is quite hard to calculate such a monotone
without knowing the structure of LHSs. Instead, we find a way to
derive lower and upper bounds based on SDPs.

\begin{figure}[t]
\centering
\includegraphics[width=1\columnwidth]{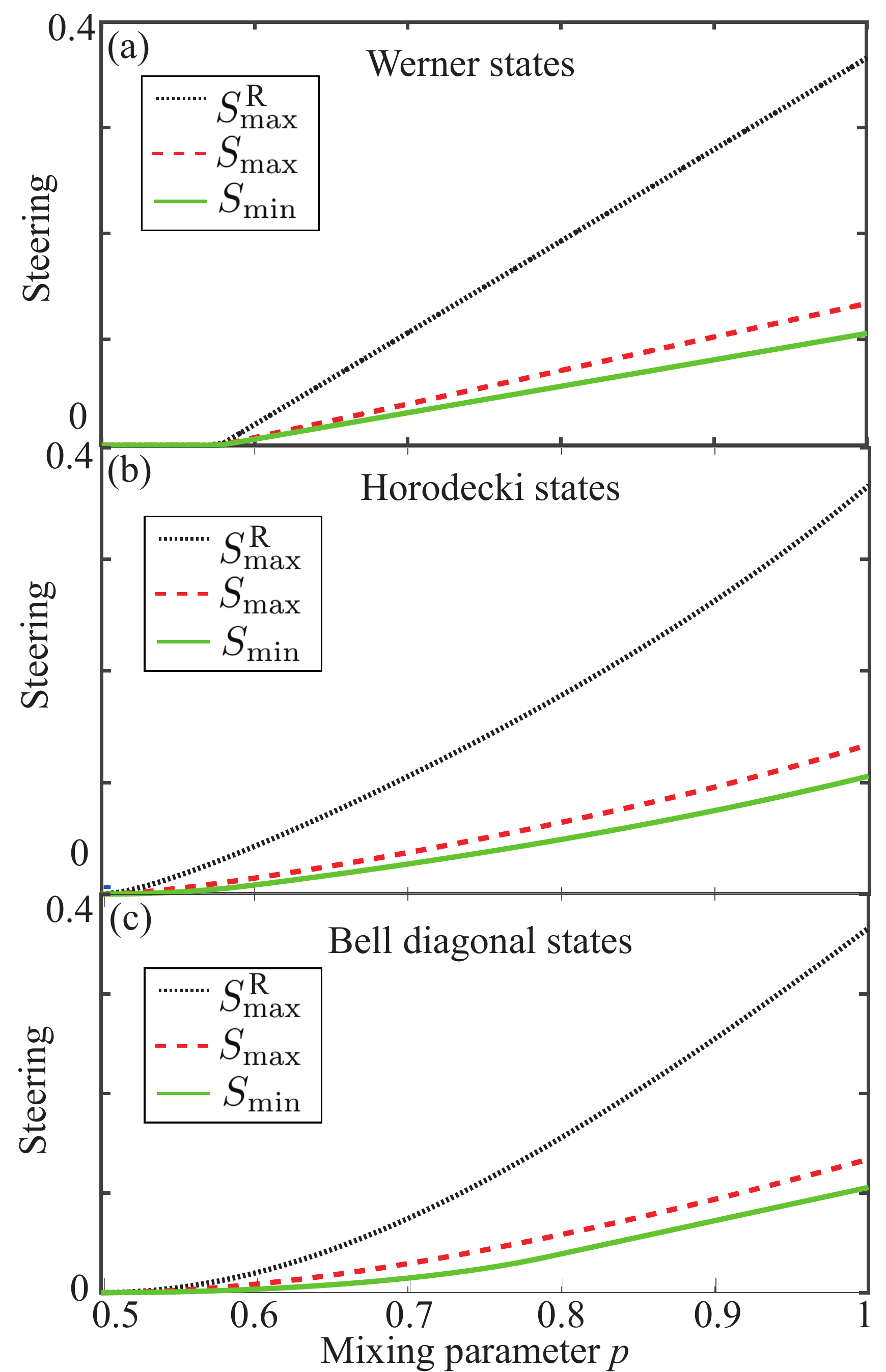}
\caption{Upper ($\Smax^{\text{R}}$ and $\Smax$) and lower
($\Smin$) bounds of the steerability measure $S_{\rm TD}$ for (a)
the Werner states (rank-4 Bell-diagonal states), (b) Horodecki
states, and (c) rank-2 Bell-diagonal states vs their mixing
parameter $p$. A given assemblage is created when Alice performs
three mutually-unbiased measurements (eigenvectors of the Pauli
spin matrices $X,~Y$, and $Z$). Here, we compute the unsteerable
assemblage obtained from the restricted-noise consistent steering
robustness (RNCSR, $S_{\rm CSR}^{\text{R}}$) bounded by
$\Smax^{\text{R}}$, and the consistent steering robustness (CSR,
$S_{\rm CSR}$) bounded by $\Smax$. As can be seen, the
steerability monotonically increases with increasing parameter,
for (b), (c) $p\ge 1/2$ and (a) $p\geq1/\sqrt{3}$. Note
that the meaning of the mixing parameter $p$  is completely
different  in (a), (b), and (c) as given by the definitions
of the corresponding states in Sec. V.A, V.B, and V.C.}
 \label{fig1}
\end{figure}

\subsection{Upper bound based on the restricted-noise consistent steering robustness}

An upper bound of $S_{\rm TD}$ can be obtained via the notion of
\emph{steering robustness}~\cite{Piani15}, i.e., the amount of
noise that can be added to an assemblage to make it unsteerable,
and the notion of CSR \cite{Cavalcanti16}, i.e., with the
requirement that the noise assemblage has the same reduced state.
We introduce a robustness measure based on this kind of mixing
with a reduced state, which can be summarized as follows: Given an
assemblage $\{ \sigma_{a|x}\}_{a,x}$ and the associated reduced
state $\sigma_B=\sum_a \sigma_{a|x}$, we define a steering
monotone as
\begin{equation}\label{eq:srg}
\begin{split}
S_{\rm CSR}^{\text{R}} (\{\sigma_{a|x}\}_{a,x}) = \min  \Big\{ t \left|\  \frac{\sigma_{a|x} + t [p(a|x) \sigma_B]}{1+t} \right.\\
\text{ is unsteerable } \Big\},
\end{split}
\end{equation}
which can be referred to as a {\it restricted-noise consistent
steering robustness} (RNCSR), where $p(a|x) = \tr(\sigma_{a|x})$.
The quantity $S_{\rm CSR}^{\text{R}}$  can be efficiently computed
as an SDP (see Appendix~\ref{app:sdp}).

Given an assemblage $\{\sigma_{a|x}\}_{a,x}$,
the unsteerable assemblage,
which is obtained as the solution of the SDP for calculating ${\rm
S}_{\rm CSR}^{\text{R}}$, is denoted by $\{\sigma^{\text{R}}_{a|x}\}_{a,x}$ . We can then easily compute the distance between
these two assemblages as follows
\begin{equation}\label{eq:Dwnsr}
\begin{split}
D_{A}(\{\sigma_{a|x}\}_{a,x},\{\sigma^{\text{R}}_{a|x}
\}_{a,x})&=
\\ \frac{1}{N_x}\frac{t_{\text{min}}
}{1+t_{\text{min}}} &
\sum_{a,x}p(a|x)D(\tilde{\sigma}_{a|x},\sigma_B),
\end{split}
\end{equation}%
where $t_{\text{min}}$ is the optimal parameter $t$ obtained from
Eq.~(\ref{eq:srg}) and the tilde denotes normalized states, e.g.,
$\tilde\sigma_{a|x}=\sigma_{a|x}/\tr(\sigma_{a|x})$. As a
consequence, the minimal trace distance between an assemblage and
the restricted set of unsteerable assemblage obtained via mixing
with noise $\sigma_{B}$, corresponds to substituting $t_{\text{min}}$
into the solution of an SDP for the $S_{\rm CSR}^{\text{R}}$ in
Eq.~\eqref{eq:Dwnsr}. Thus, an upper bound on $S_{\rm TD}$ can
be simply given by
\begin{equation}
S_\max^{\text{R}}(\{\sigma_{a|x}\}_{a,x} ) =
D_{A}(\{\sigma_{a|x}\}_{a,x},\{\sigma^{\text{R}}_{a|x}\}_{a,x}).
\label{eq:SmaxGamma}
\end{equation}
Note that if Bob's system is a qubit, then
$S^{\text{R}}_{\max}(\{\sigma_{a|x}\}_{a,x})$ corresponds to a
half of the sum of all the Euclidean distances between the Bloch
vectors  $\tilde{\sigma}_{a|x}$ and $\sigma_{B}$ in the Bloch
sphere multiplied by the probability distribution $p(a|x)$ and the
scaling factor $t_\min/[N_x(1+t_\min)]$. Mathematically, this can
be expressed as
\begin{equation}
S^{\rm{R}}_{\rm{max}}(\{\sigma_{a|x}\}_{a,x})=\sum_{a,x}\frac{p(a|x)\,
t_{\min}}{N_{x}(1+t_{\min})}\frac{|\vec{p}_{a|x}-\vec{q_{b}}|}{2},
\end{equation}
where $|\vec{a}-\vec{b}|$ denotes the Euclidean distance between
vectors $\vec{a}$ and $\vec{b}$. Moreover,
$\vec{p}_{a|x,i}=\text{tr}(\tilde{\sigma}_{a|x}\sigma_{i})$ and
$\vec{q}_{b,i}=\text{tr}(\sigma_{B}\sigma_{i})$ (for $i=1,2,3$)
are the components of the Bloch vectors of $\tilde{\sigma}_{a|x}$
and $\sigma_{B}$, respectively, and
$\{\sigma_1,\sigma_2,\sigma_3\}\equiv\{X,Y,Z\}$ denote the Pauli
operators.

\subsection{Upper bound based on the consistent steering robustness}

Here we provide another upper bound on the steering monotone $S_{\rm TD}$,
which is also based on the CSR. We show that this new
bound is even tighter than that of
$\Smax^{\text{R}}(\{\sigma_{a|x}\}_{a,x} )$, as defined in
Eq.~(\ref{eq:SmaxGamma}).

The CSR is defined as follows~\cite{Cavalcanti16}:
\begin{equation}\label{eq:sr}
\begin{split}
S_{\rm CSR} (\{\sigma_{a|x}\}_{a,x}) = \min  \Big\{ t \left|\  \frac{\sigma_{a|x} + t \tau_{a|x}}{1+t} \right.\\
\text{ is unsteerable, and } \sum_a \tau_{a|x}=\sigma_B,\ \forall x   \Big\},
\end{split}
\end{equation}
where $\{\tau_{a|x}\}_{a,x}$ is an arbitrary noise assemblage with the same reduced state.

Similarly to $\Smax^{\text{R}}$, an upper bound of $S_{\rm TD}$ can be
obtained from the optimal solution
$\{\sigma^{\text{CSR}}_{a|x}\}_{a,x}$ of an SDP for the CSR. Note
that although the $\{\sigma^{\text{CSR}}_{a|x}\}_{a,x}$ is an
optimal unsteerable assemblage, it may not be closest according to the trace distance.
Thus, an upper bound based on the CSR can be defined as
\begin{equation}\label{eq:Dsr}
\begin{split}
\Smax(\{\sigma_{a|x}\}_{a,x})=D_{A}(\{\sigma_{a|x}\}_{a,x},\{\sigma^{\text{CSR}}_{a|x}
\}_{a,x}).
\end{split}
\end{equation}%
Because $\Smax^{\text{R}}$ was obtained for a restricted noise, so
it is obvious that the following inequality holds in general:
\begin{equation}
  \Smax\leq  \Smax^{\text{R}}.
 \label{compare}
\end{equation}

\subsection{Lower bound based on operator norm}

In this section, we show how to compute a lower bound of $S_{\rm TD}$ as an
SDP.
Without loss of generality, we can write the assemblage
$\{\rho_{a|x}\}_{a,x}\in \text{LHS}$, as $\rho_{a|x} =
\sum_\lambda \delta_{a,\lambda_x} \sigma_\lambda$, where $\lambda$
is a vector $(\lambda_x)_x$ and $\delta_{a,\lambda_x}$ represent
the deterministic strategy for choosing the assemblage element
$\sigma_\lambda$~\cite{Pusey13}. For consistency, we assume the condition $\sum_\lambda \sigma_\lambda = \sum_a \rho_{a|x}=\sigma_B$.
Then we note that the trace norm
can be lower bounded by the operator norm
\begin{equation}
  \| A \|_\infty := \min \{\mu | -\mu \openone \leq
A \leq \mu \openone\},
 \label{N107}
\end{equation}
i.e., $\| A \|_\infty \leq \| A\|$ for all operators $A$.
Combining the lower bound based on this norm with the definition
of the unsteerable assemblage, we obtain the following lower bound
for $S_{\rm TD}(\{\sigma_{a|x}\}_{a,x} )$:
\begin{equation}\label{eq:SDP_int}
\begin{split} \Smin := \min_{\substack{\sigma_\lambda}}: & \frac{1}{2N_x}\sum_{a,x} \| \sigma _{a|x} - \sum_\lambda \delta_{a,\lambda_x} \sigma_\lambda \|_\infty \\
\text{subject to: }& \sum_\lambda \sigma_\lambda = \sigma_B;\\
   & \sigma_\lambda \geq 0, \quad \forall \lambda;
\end{split}
\end{equation}
where $\delta_{a,\lambda_x}$ is the usual deterministic strategy
for the LHS model. Now, we can rewrite the above problem as the
following SDP:
\begin{equation}\label{eq:Smin}
\begin{split}
\Smin = \min_{\substack{\mu_{a,x},\sigma_\lambda}}: & \frac{1}{2N_x}\sum_{a,x} \mu_{a,x} \\
\text{subject to: }& -\mu_{a,x} \openone \leq \sigma _{a|x} - \sum_\lambda \delta_{a,\lambda_x} \sigma_\lambda \leq \mu_{a,x} \openone\ \\
    & \sum_\lambda \sigma_\lambda = \sigma_B;\\
   & \sigma_\lambda \geq 0, \quad \forall \lambda.
\end{split}
\end{equation}
By definition, we have $0 \leq \mu_{a,x}\leq 1$, so the same holds
for the solution of the SDP. This implies that the primal SDP
problem is bounded. Moreover, it is also strictly feasible, e.g.,
just take any  strictly positive assemblage $\sigma_\lambda$, consistent with the reduced state $\sigma_B$,
and $\mu_{a,x}=1$ for all $a,x$. This implies the strong duality
condition, i.e., the primal and dual SDPs have the same optimal
value.

Note, however, that the operator norm quantifier $S_{\rm
min}(\{\sigma_{a|x}\}_{a,x})$ is \emph{not} a convex steering
monotone, since the operator-norm distance is, in general, not
contractive under completely positive trace-nonincreasing maps.

Finally, we have the following lower and upper bounds:
\begin{equation}
\begin{split}
\Smin(\{\sigma_{a|x}\}_{a,x} ) \leq
 S_{\rm TD}(\{\sigma_{a|x}\}_{a,x} ) \leq \Smax(\{\sigma_{a|x}\}_{a,x})\\
\leq \Smax^{\text{R}}(\{\sigma_{a|x}\}_{a,x} ),
\end{split}
\end{equation}
which can be efficiently computed via our SDPs.
A clear comparison of these three upper bounds and lower bound for some states is
shown in Fig.~\ref{fig1}.

\section{Geometric witness of steerability}\label{sec:witness}

In Sec.~III, we concentrated on assemblages, but in this section we
focus on steerability of a quantum state rather than assemblages.

In addition to the geometrical picture introduced in Sec.~III, we
provide a way to visualize two-qubit steering properties through
the notion of a LHS surface and a QSE. We first recall that the
QSE~\cite{Jevtic14} is defined as the surface of normalized
assemblages
$\tilde{\sigma}_{a|x}^{{}}=\sigma_{a|x}^{{}}/\text{tr}(\sigma
_{a|x}^{{}})$, obtained by Bob for all possible projective
measurements of Alice. All projective measurements on Alice's side
form the surface of the QSE, while the POVMs correspond to the
points in the interior. The QSE centers at
$\tilde{c}=(\tilde{b}-T^{\text{T}}\tilde{a})/(1-\tilde{a}^{2})$
with the orientation and semiaxes lengths $s_{i}=\sqrt{q_{i}}$
given by the eigenvectors and eigenvalues $q_{i}$ of the ellipsoid
matrix
\begin{equation}
Q=\frac{1}{1-\tilde{a}^{2}}(T^{\text{T}}-\tilde{b}
\tilde{a}^{\text{T}})\left(\openone+\frac{\tilde{a}
\tilde{a}^{\text{T}}}{1-\tilde{a}^{2}}\right)(T-\tilde{a}\tilde{b}^{\text{T}}),
\end{equation}%
where $\tilde{a}$ and $\tilde{b}$ are the Bloch vectors of the
reduced states of Alice and Bob, respectively. Here, $T$ is the
correlation matrix with elements
$T_{jk}=\text{tr}[\rho_{\text{AB}} \sigma _{j}\otimes \sigma
_{k}]$ (for $j,k=1,2,3$), where $\rho_{\text{AB}}$ is the
bipartite state shared by Alice and Bob.

We can analogously define the corresponding LHS surface. Instead
of considering all possible single measurements, however, we need
to fix a measurement assemblage for Alice. In this case, we assume
that Alice can perform three mutually unbiased
measurements~\footnote{We need to fix the number of measurements
on Alice's side to be able to define some steerability
properties.} on her side with outcomes $\pm 1$. Consequently, Bob
obtains the assemblage $\{\sigma _{a|x}\}_{a,x}$, consisting of
six terms. To compute the closest unsteerable assemblage,
$\{\sigma _{a|x}^{\text{US}}\}_{a,x}$, we restrict to the RNCSR
case which can be computed as an SDP. By normalizing such an
assemblage, $\tilde{\sigma}_{a|x}^{\text{US}}=\sigma
_{a|x}^{\text{US}}/\text{tr}(\sigma _{a|x}^{\text{US}})$, Bob
obtains six vectors in the Bloch sphere. The LHS surface is then
obtained when Alice performs all possible rotations of her
mutually-unbiased measurement bases.

Intuitively, the bipartite state $\rho_{AB}$ is unsteerable if its
LHS surface and QSE are identical because
$\{\tilde{\sigma}_{a|x}^{{}}\}_{a,x}=\{\tilde{\sigma}_{a|x}^{\text{US}}\}_{a,x}$.
Moreover, it is clear that  ${\rm conv}($LHS surface$)$ is always
contained in ${\rm conv}($QSE$)$, where we denoted with ${\rm
conv}$ the convex hull of the points in the corresponding surface.
In fact, given $\{\sigma _{a|x}\}_{a,x}$, the corresponding
solution $\{ \sigma_{a|x}^{\text{US}} \}_{a,x}$, computed via an
SDP for the RNCSR, satisfies $\tr(\sigma_{a|x}) =
\tr(\sigma_{a|x}^{\text{US}})$, and hence
$\tilde{\sigma}_{a|x}^{\text{US}}$ is a convex hull of points
inside the QSE, namely,
\begin{equation}
\tilde{\sigma}_{a|x}^{\text{US}}= \frac{\tilde{\sigma}_{a|x} + t
\sigma_{B}}{1+t}.
\end{equation}
Therefore, we can geometrically witness steering when
\begin{equation}\label{eq:volume}
\Delta V\equiv V_{\text{QSE}}-V_{\text{LHS surface}} > 0,
\end{equation}%
where $V_{\text{QSE}}$ and $V_{\text{LHS surface}}$ are the volumes
of the QSE and LHS surface, respectively.

Note that the steering witness $\Delta V$ focuses on the
steerability of a \emph{quantum state}, while the proposed
steering monotone $S_{\rm TD}$ describes the steerability of an
\emph{assemblage}. Thus, one could think that it is rather hard,
in general, to compare these approaches and to show which of these
is more useful. Now we would like to explain now an important
relation between the witness $\Delta V$ and the steering monotone
$S_{\rm TD}$. Note that $S_{\rm TD}$ is defined on an assemblage,
hence, it requires the measurement settings to be fixed. In
contrast to this, the calculation of $\Delta V$ involves looking
at all possible mutually-unbiased bases; hence, it provides a more
complete information about the steerability of a given state. In
addition, McCloskey {\it et al.}~\cite{McCloskey16} showed that
the geometric information encoded in the QSE often provides the
optimal measurement directions, corresponding to the three
ellipsoid semi-axes. Similarly, the LHS surface provides the
information about the measurement directions giving usually the
highest steering monotone $S_{\rm TD}$.

The concept of the LHS surface can be generalized to include
different SDP characterizations of the ``closest'' unsteerable
assemblages, e.g., via the steering robustness~\cite{Piani15} or
other quantifiers~\cite{SDPreview17}. Moreover, such a notion can
also be generalized beyond the qubit case. The interest for the
present approach is motivated by the possibility of visualizing
the steering properties of a state onto the Bloch sphere and its
relations with the QSE.

\section{Applications}\label{sec:appli}

In Sec.~\ref{sec:quanti}, we provided examples of lower and upper
bounds for our steering monotone $S_{\rm TD}$. In what follows, we
demonstrate the usefulness of the LHS surface and the
trace-distance measures of steerability in several related
examples.

We analyze three important prototype classes of states (i.e., the
Werner, Horodecki, and Bell-diagonal states), which are formed by
the singlet state $|S\rangle$ mixed with three different states.
Thus, the meaning of the mixing parameter is different in these
states although denoted, for simplicity, by the same symbol $p$.
Specifically, (a) for the Werner states, the singlet state
$|S\rangle$ is mixed with a (separable) completely mixed state,
which is not orthogonal to $|S\rangle$; (b) for the Horodecki
states, $|S\rangle$ is mixed with a separable state, which is
orthogonal to $|S\rangle$; and (c) for the Bell-diagonal states,
$|S\rangle$ is mixed with another maximally entangled state (i.e.,
the entangled triplet state), which is orthogonal to $|S\rangle$.

\begin{figure*}[t!]
\centering \includegraphics[width=2\columnwidth]{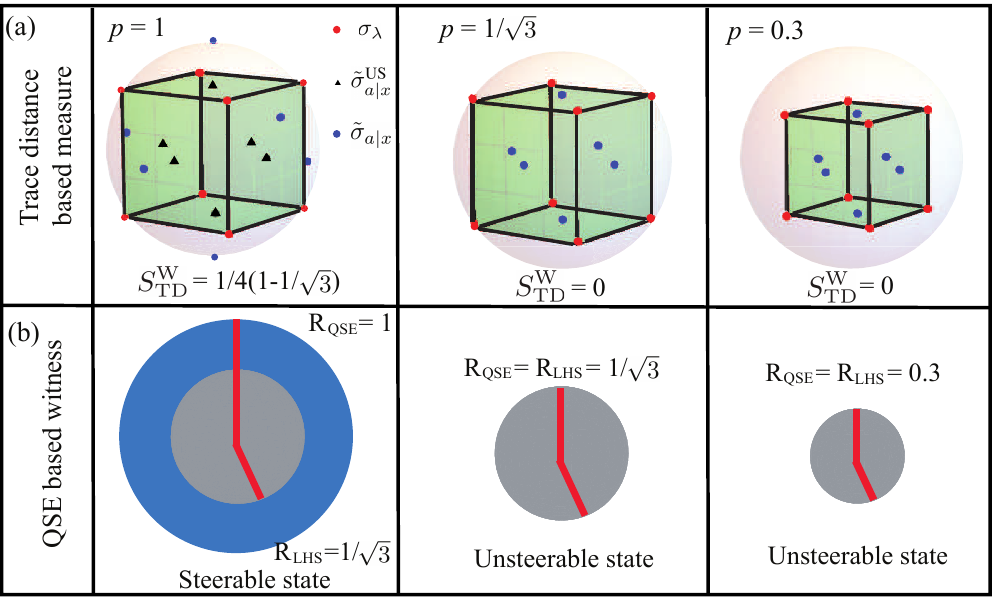}
\caption{EPR steerability of the Werner states: (a) steerability
measure $S_{\rm TD}^{\rm W}$, which is based on trace distance,
and (b) steerability witness given, by Eq.~(\ref{eq:volume}),
which is based on the volume of the QSE and LHS surface. For
(a), we use the three Pauli bases to obtain the preexisted states
$\{\protect\sigma _{\protect\lambda }\}_{ \protect\lambda }$
(red circles), steered states $\{\tilde{\protect\sigma} _{a|x}\}_{a,x}$
(black triangles), and the unsteerable assemblage
$\{\tilde{\protect\sigma} _{a|x}^{\text{US}}\}_{a,x}$ (blue circles),
respectively. The auxiliary green cube helps one to see the relative
positions of the LHS and the steered states. As expected, when
$p\geq $ $1/\protect\sqrt{3}$, we find that the positions of the
steered states are outside the LHSs. On the other hand, the Bloch
vectors of $\{\tilde{\protect\sigma}_{a|x}^{\text{US}}\}_{a,x}$
remain as ($\pm 1/\protect\sqrt{3}$,0,0), (0,$\pm
1/\protect\sqrt{3}$,0) and (0,0,$\pm 1/\protect\sqrt{3}$),
independent of $p$. This is because the maximal length of a Bloch
vector is equal to unity. When $p\leq 1/\protect\sqrt{3}$, the
unsteerable states
$\{\tilde{\protect\sigma}_{a|x}^{\text{US}}\}_{a,x}$ are exactly
identical to the steered states
$\{\tilde{\protect\sigma}_{a|x}\}_{a,x}$. For (b), we show the
QSE and LHS surface for different values of $p$. Both surfaces are
spheres, so it is sufficient to show the two-dimensional
projection. The outer and inner circles are the QSE and LHS
surface, respectively. These circles are centered at
$\tilde{c}=(0,0,0)$. The three semiaxes of the QSE lie along $x$,
$y$, and $z$. However, the orientations of the three LHS axes are
the same as those of the QSE, but the length is $p$ only when
$0<p<1/\protect\sqrt{3}$. When $p>1/\protect\sqrt{3}$, the lengths
are fixed at $1/\protect\sqrt{3}$. In other words, the QSE and LHS
surfaces are identical when $p\leq 1/\protect\sqrt{3}$. Once
$p\geq 1/\protect\sqrt{3}$, the LHS surface is fixed, but the QSE
expands with $p$. } \label{fig2}
\end{figure*}

\begin{figure*}[tbp]
\includegraphics[width=1.8\columnwidth]{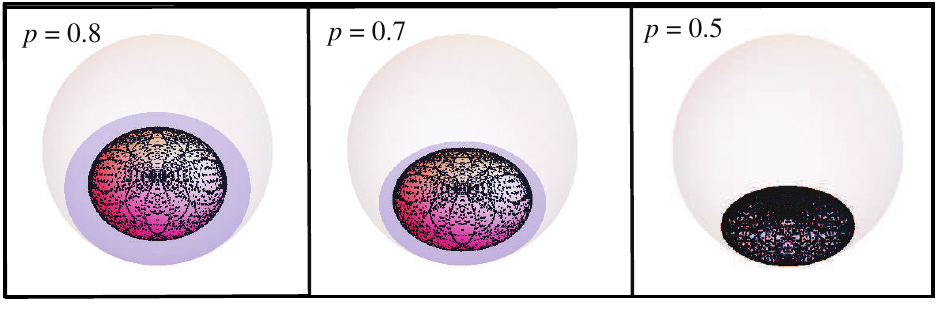}
\caption{QSE-based witness of the steerability of the Horodecki
states. As can be seen, a given state is steerable if the LHS
surface and QSE are not identical. Horodecki states are
unsteerable if the parameter $p=1/2$. Otherwise these states are
steerable.  Note that one of the lower bounds of $S_{\rm{TD}}$
corresponds to the distance between the inner and outer surfaces
(i.e., the Euclidean distance between
$\tilde{\sigma}_{a|x}^{\text{US}}$ and $\tilde{\sigma}_{a|x}$ in
the Bloch sphere) multiplied by the probability distribution
$p(a|x)$ and some scaling factor defined in the text. }
\label{fig3}\centering %
\end{figure*}

\begin{figure*}[tbp]
\includegraphics[width=1.8\columnwidth]{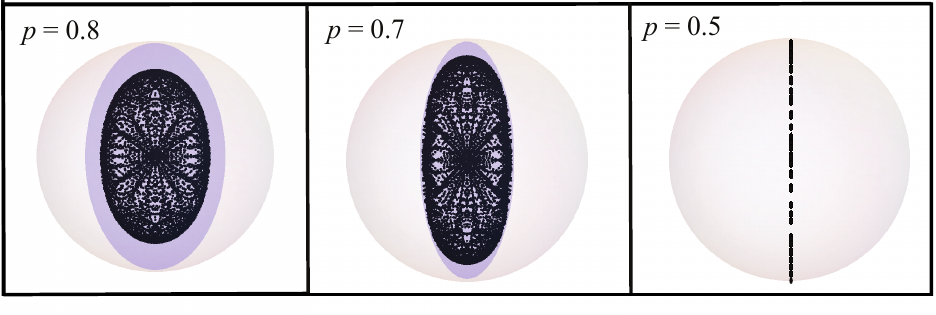}
\caption{QSE-based witness of the steerability of the rank-2
Bell-diagonal states. As can be seen, a given state is steerable
if its LHS surface and QSE are not identical. The Bell-diagonal
states are unsteerable if the parameter $p=1/2$. Otherwise these
are steerable.}
\label{fig4}\centering %
\end{figure*}

\subsection{Steerability of Werner states}

We analytically show the solution of the LHS surface for the
Werner states~\cite{Werner89}, which are mixtures of the
singlet state and the maximally-mixed state, i.e.,
\begin{equation}
  {\rho^{\rm W}(p)=p|S\rangle \langle S|+(1-p)\frac{\openone}{4}},
 \label{N104}
\end{equation}
where ${|S\rangle =1/\sqrt{2}(|10\rangle -|01\rangle )}$ and
$0\leq p\leq 1$ is the mixing weight. It is clear that the Werner
states are rank-4 Bell-diagonal states for $p<1$. When Alice
applies the three Pauli operators ($X$, $Y$, and $Z$) to measure a
given Werner state, the corresponding Bloch vectors of Bob's
normalized assemblage are ($\pm p$,0,0), (0,$\pm p$,0), and
(0,0,$\pm p$). The simplest solution of the preexisted quantum
states $\{\sigma _{\lambda }\}_{\lambda }$ is located at ($\pm
p$,$\pm p$,$\pm p$) in the Bloch sphere. On the other hand, the
LHSs are the mixtures of four preexisted states according to the
strategy $p(\lambda )$ with probability $p(a|x,\lambda )=1/4$.
When $p\leq 1/\sqrt{3}$, the LHSs
$\tilde{\sigma}_{a|x}^{\text{US}}$ are identical to the steered
states $\tilde{\sigma}_{a|x}$. As $p\geq $ $1/\sqrt{3}$, the Bloch
vectors of $\tilde{\sigma}_{a|x}^{\text{US}}$ are fixed at ($\pm
1/\sqrt{3}$,0,0), (0,$\pm 1/\sqrt{3}$,0), and (0,0,$\pm
1/\sqrt{3}$) as shown in Fig.~\ref{fig2}. This is because the
maximal length of a Bloch vector is equal to unity. The optimal
value of $p$ for the LHS and preexisted states is $1/ \sqrt{3}$,
coinciding with the upper bound $S_{\max}^{\text{R}}$ on the
steering inequality $\langle XX\rangle +\langle YY\rangle +\langle
ZZ\rangle \leq \sqrt{3}$~\cite{Cavalcanti09}. However, the set of
steered states $\tilde{\sigma }_{a|x}$, i.e., the QSE, gradually
expands with $p$. One can also rotate the measurement settings on
Alice's side, but keep them mutually unbiased. Once all sets of
the three measurements are performed, Bob obtains the LHS surface,
which is the set of all $\tilde{\sigma}_{a|x}^{\text{US}}$~(see
Fig.~\ref{fig2}). One can solve analytically this simple case and
show that the LHS surface of the Werner state is actually a
sphere, centered at $\tilde{c}=(0,0,0)$ as the QSE, with radius
$1/\sqrt{3}$ for $p \geq 1/\sqrt{3}$, and radius $p$ otherwise
(see Fig.~\ref{fig2} and Appendix~D). The trace distance between
them is equal to
\begin{equation}
D_A^{\rm W}(\{\sigma_{a|x}\}_{a,x})= \frac{1}{4}\left(p-\frac{1}{\sqrt{3}}\right)
\end{equation}
for $p\geq 1/\sqrt{3}$, and $0$ otherwise, which is identical to
quarter the Euclidean distance between the
$\tilde{\sigma}_{a|x}^{\text{ US}}$ and $\tilde{\sigma}_{a|x}$.
Interestingly, the $D_A^{\rm W}(\{\sigma_{a|x}\}_{a,x})$, which
can be computed by our analytical solution for the Werner states,
is smaller than $S_{\text{max}}$ and the same as $S_{\text{min}}$.
Thus, we conclude that $S^{\rm{W}}_{\text{TD}}$=$D_A^{\rm
W}(\{\sigma_{a|x}\}_{a,x})$, where the superscript $W$ indicates
that the results are for the Werner states. Note that in this
example, we considered only three Pauli measurement bases. It can
be constructive to compare Fig.~\ref{fig2} for the Werner states
with Figs.~\ref{fig3} and~\ref{fig4} for other special states.

Another comparison of various upper and lower bounds for the
Werner states is shown in Fig.~\ref{fig1}(a). It is seen that the
upper bounds of $S_{\rm TD}^{\rm{W}}$ for the Werner states are
vanishing for the mixing parameter $p\le1/\sqrt{3}$ and are
linearly increasing with $p>1/\sqrt{3}$. This is contrary to the
behavior of the same bounds for the states analyzed in
Figs.~\ref{fig1}(b) and~\ref{fig1}(c).

\subsection{Steerability of Horodecki states}

The Horodecki states are the mixtures of a maximally entangled
state, say the singlet state $|S\rangle $, and a separable state,
say $|00\rangle $, i.e.,
\begin{equation}
  \rho ^{\rm H}(p)=p|S\rangle \langle S|+(1-p)|00\rangle \langle 00|.
 \label{N105}
\end{equation}
In Fig.~\ref{fig3}, we show that the LHS surfaces, which are
computed by the RNCSR for the Horodecki states, are similar to
those for the Werner states. When $0\leq p\leq 1/2$, the LHS
surface and QSE are identical. Therefore, the trace distance
between a given assemblage and unsteerable assemblage, which we
consider $X,~Y,$ and $Z$, is $0$, when $0\leq p\leq 1/2$. As
$p\geq 1/2$, the QSE and LHS surfaces gradually expand but the QSE
expands more rapidly than the LHS surface. The trace distance of
the assemblages also increases when $1/2\leq p\leq 1$ [see
Figs.~\ref{fig1}(b)]. Via numerical fitting of the computed
points, we find that the LHS surface associated with the Horodecki
states is consistent with the corresponding QSE. Another
comparison of various upper and lower bounds for the Horodecki
states is shown in Fig.~\ref{fig1}(b).

\subsection{Steerability of rank-2 Bell-diagonal states}

In general, Bell-diagonal states of two qubits are mixtures of the
four maximally-entangled quantum states. Here, for simplicity, we
consider special rank-2 Bell-diagonal states, i.e., mixtures of
the singlet state $|S\rangle$ and a triplet state $|T\rangle
=1/\sqrt{2}(|10\rangle +|01\rangle )$, i.e.,
\begin{equation}
  \rho ^{\rm B}(p)=p|S\rangle \langle S|+(1-p)|T\rangle \langle
  T|,
 \label{N106}
\end{equation}
where $p$ is the mixing weight. In Fig.~\ref{fig4}, we show that
the LHS surface, which was computed by the RNCSR, is identical to
the QSE only when $p=1/2$; otherwise, these are different. The
distance between a given assemblage and unsteerable assemblage,
which we consider $X,~Y,$ and $Z$, also reveals the same behavior,
as shown in Fig.~\ref{fig1}(c). However, the LHS surface of these
Bell-diagonal states cannot be fitted by an ellipsoid.

The upper bounds of $S_{\rm TD}$ for the Horodecki states [as
shown in Fig.~\ref{fig1}(b)] and the Bell-diagonal states
[Fig.~\ref{fig1}(c)] are vanishing for the mixing parameter
$p\in[0,1/2]$ and $p=1/2$, respectively. Moreover, these bounds
are \emph{nonlinearly} increasing with $p\ge 1/2$. This is in
contrast to those upper bounds for the Werner states [shown in
Fig.~\ref{fig1}(a)], which vanish for $p\in[0,1/\sqrt{3}]$ and are
\emph{linearly} increasing with $p>1/\sqrt{3}$. As already
mentioned in the introduction to Sec. V, the meaning of the mixing
parameter $p$ for these three classes of states is completely
different. By analyzing the Werner states, we see that, by mixing
the singlet state $|S\rangle$ with the maximally mixed state
$\openone/4$, the steerability of such ``noisy'' singlet states is
completely destroyed for a wider range of the values of the mixing
parameter $p$ in comparison to the other two cases, i.e., to the
mixing of the singlet state $|S\rangle$ with a state
$|\psi\rangle$ orthogonal to $|S\rangle$ both for the separable
state $|\psi\rangle=|00\rangle$ (which results in the Horodecki
states) and for the maximally entangled state
$|\psi\rangle=|T\rangle$ (which results in the Bell-diagonal
states).

\section{Conclusions and outlook}\label{sec:conclu}

In this work, we defined the trace-distance between two
assemblages and the corresponding measure of steerability based on
this distance. We have shown that this measure of steerability is
indeed a \emph{convex steering monotone} under restricted
one-way LOCCs. We provided a way of estimating such a quantity via
lower and upper bounds based on SDPs. Specifically, a lower bound
is based on the operator norm, while a few upper bounds are
found by applying various steering measures, including
the CSR~\cite{Cavalcanti16}, and a
restricted version of the CSR.
Using the latter bound, we proposed a way of visualizing the
steerability property of a quantum state in the Bloch sphere via
the notion of a LHS surface, which relates the
steerability problem, in the sense of the existence of a
LHS model, with the notion of QSEs.

We computed EPR steerability by describing a set of states in the
Bloch sphere. We did not construct a space of assemblages. Thus,
we defined, in particular, the upper bound $S_{\max}^{\text{R}}$,
which can be directly computed by summing up (with some
coefficients) all the ``Euclidean distances'' between Bloch
vectors. Therefore, $S_{\max}^{\text{R}}$ has a clear geometrical
meaning. Moreover, in Sec.~V.A, we also pointed out that $S_{\rm
TD}$ for the Werner states (as denoted by $S_{\rm TD}^{\rm W}$)
has a direct relation to the distance between the set of states in
the Bloch sphere. The monotone $S_{\rm TD}^{\rm W}$ is a linear
function of the distance from the points $(x,0,0)$, $(0,x,0)$, and
$(0,0,x)$ (where $x=\pm 1/\sqrt{3}$ ) of the Bloch sphere to a
normalized quantum state of its assemblage in the Bloch sphere,
when the mixing parameter $p>1/\sqrt{3}$. Thus, by referring to a
``geometrical'' interpretation, we mean the Euclidean distance
between quantum states in the Bloch sphere.

We defined the witness $\Delta V$ of steerability corresponding to
the difference of the volumes enclosed by the QSE and the LHS
surface. The LHS surfaces enable calculation of the proposed steering
monotone $S_{\rm TD}$ optimized over all mutually-unbiased bases.
We remark that this observation relates the different concepts of
(i) the steering witness $\Delta V$, which reveals the
steerability of a \emph{quantum state}, and (ii) the steering
monotone $S_{\rm TD}$, which describes the steerability of an
\emph{assemblage}.

Our study stimulates some further investigations. First, it is
known that the QSE has an analytical
representation. Therefore, it is natural to ask if the LHS surface
also has an analytical formulation, at least, for some specific
states. Our analysis shows that this is the case for the Werner
states and we have numerical evidence that the same could be possible
for the Horodecki states. Second, since we have obtained the witness of
steerability for a given quantum state, given in
Eq.~\eqref{eq:volume}, it is interesting to investigate whether
such a difference of volumes, i.e., between the QSE and LHS
surface, has some physical meaning and can be used to obtain a
new steering monotone for a quantum state.

\begin{acknowledgments}
We thank Hong-Bin Chen, Che-Ming Li, Yeong-Cherng Liang, and Mark
M. Wilde for helpful discussions. In particular, we thank Mark
Wilde for his clarifications on the notion of one-way LOCC
steering monotone and its restricted version, and for finding a
related error in an earlier version of the manuscript. This work
was supported partially by the National Center for Theoretical
Sciences and Ministry of Science and Technology, Taiwan, under Grant No. MOST 103-2112-M-006-017-MY4. C.B. acknowledges support
from FWF (Project: M~2107 Meitner-Programme). Y.N.C., A.M., and
F.N. acknowledge the support of a grant from the Sir John
Templeton Foundation. F.N. is partially supported by the MURI
Center for Dynamic Magneto-Optics via the AFOSR Award No.
FA9550-14-1-0040, the Japan Society for the Promotion of Science
(KAKENHI), the IMPACT program of JST, CREST Grant No. JPMJCR1676,
RIKEN-AIST Challenge Research Fund, and JSPS-RFBR Grant No.
17-52-50023.
\end{acknowledgments}

\appendix
\section{Metric properties}\label{app:metric}

Here we define the trace distance between two assemblages as
\begin{equation}
D_{A}(\{\sigma_{a|x}\}_{a,x},\{\rho_{a|x}\}_{a,x})=\frac{1}{N_x}\sum_{a,x}D(\sigma_{a|x},\rho_{a|x}),
\end{equation}
where $D(\rho,\rho'):=\frac{1}{2}\|\rho-\rho'\|$ and
$\|X\|:=\text{tr}[|X|]$ is the trace norm.

Now we show that $D_A$ satisfies the following three basic properties required for a true metric:
(1) It is obvious that
$D_{A}(\{{\sigma_{a|x}}\}_{a,x},\{{\rho_{a|x}}\}_{a,x})=0$ because
$\{\sigma_{a,x}\}_{a|x}$ and $\{\rho_{a|x}\}_{a,x}$ are the same.

(2) Here, we prove that the trace distance between two assemblages is symmetric:
\begin{equation}
\begin{aligned}
D_{A}(\{{\sigma_{a|x}}\}_{a,x},\{{\rho_{a|x}}\}_{a,x})&=\frac{1}{N_x}\sum_{a,x}D(\{\sigma_{a|x}\}_{a,x},\{\rho_{a|x}\}_{a,x})\\&=\frac{1}{N_x}\sum_{a,x}D(\{\rho_{a|x}\}_{a,x},\{\sigma_{a|x}\}_{a,x})\\&=D_{A}(\{{\rho_{a|x}}\}_{a,x},\{{\sigma_{a|x}}\}_{a,x}).
\end{aligned}
\end{equation}
The second equality is based on a property of the matrix norm.

(3) We now show that the trace distance between two assemblages satisfies the triangle
inequality,
\begin{equation}
\begin{aligned}
&D_{A}(\{{\sigma_{a|x}}\}_{a,x},\{{\rho_{a|x}}\}_{a,x})=\frac{1}{N_x}\sum_{a,x}D(\{\sigma_{a|x}\}_{a,x},\{\rho_{a|x}\}_{a,x})\\&\leq
\frac{1}{N_x}\sum_{a,x}\left[D(\{\sigma_{a|x}\}_{a,x},\{\theta_{a|x}\}_{a,x})
+D(\{\theta_{a|x}\}_{a,x},\{\rho_{a|x}\}_{a,x})\right]
\\&=D_{A}(\{{\sigma_{a|x}}\}_{a,x},\{{\theta_{a|x}}\}_{a,x})+D_{A}(\{{\theta_{a|x}}\}_{a,x},\{{\rho_{a|x}}\}_{a,x}).
\end{aligned}
\end{equation}
The first inequality follows from the property of the trace
norm. This completes our proof.

\section{Restricted convex steering monotone}\label{app:mono}

First, we recall the definition of a convex steering monotone
introduced in Ref.~\cite{Gallego15} with the restrictive assumption from Ref.~\cite{Eneet17a}, namely, the independence of Alice's choice from Bob's outcome $\omega$.

A function $S$, relating
assemblages with non-negative real numbers, is a convex steering
monotone if it satisfies the following:
\begin{itemize}
\item[(i)] It vanishes for unsteerable assemblages:
\begin{equation}
S(\{\sigma_{a|x}\}_{a,x})=0\quad \text{for~all}\quad
\{\sigma_{a|x}\}_{a,x}\in\text{LHS}.
\end{equation}
\item[(ii)] (Monotonicity) $S$ is non-increasing, on average, under restricted one-way LOCCs, i.e.,
\begin{equation}
\sum_\omega P(\omega)\ S\left(\left\{ \frac{\sigma_{a'|x'}^\omega }{P(\omega)} \right\}_{a',x'} \right)\leq S(\{\sigma_{a|x}\}_{a,x}),
\end{equation}
where
\begin{equation}\label{eq:wire_rest}
\sigma^{\omega}_{a'|x'}:=\sum_{a,x}p(x|x')p(a'|a,x,x',\omega)K_{\omega}\sigma_{a|x} K_{\omega}^\dagger
\end{equation}
is an assemblage obtained from the initial assemblage $\{\sigma_{a|x}\}_{a,x}$
by performing restricted one-way LOCCs. Here, $K_\omega$ is a Kraus operator with outcome $\omega$ and $\sum_{i}K_{i}^\dagger K_i=\openone$, while $p(x|x')$ and $p(a'|a,x,x',\omega)$ are classical postprocessing, i.e., deterministic wiring maps, on Alice's side.
\item [(iii)](Convexity) Given a real number $0\leq \mu \leq 1$, and two assemblages
$\{\sigma_{a|x}\}_{a,x}$ and $\{\sigma'_{a|x}\}_{a,x}$, the
steering function $S$ satisfies the inequality
\begin{equation}
\begin{split}
S(\mu\{\sigma_{a|x}\}_{a,x}+(1-\mu)\{\sigma'_{a|x}\}_{a,x})\hspace{2cm}\nonumber\\
\leq \mu S(\{\sigma_{a|x}\}_{a,x})
+(1-\mu)S(\{\sigma'_{a|x}\}_{a,x}).
\end{split}
\end{equation}
\end{itemize}

Given an assemblage, we recall that the consistent trace-distance measure of
steerability is defined as
\begin{equation}
\begin{split}
S_{\rm TD}(\{\sigma_{a|x}\}_{a,x} ):=\min \Big\lbrace D_{A}({\{\sigma_{a|x}\}_{a,x},\{\rho_{a|x}\}_{a,x}}) \Big| \\
{\{\rho_{a|x}\}_{a,x}\in \text{LHS}}, \  \sum_a \rho_{a|x} = \sum_a \sigma_{a|x}, \ \forall x\ \Big\rbrace.
\end{split}
\end{equation}
where
$D_{A}(\{\sigma _{a|x}\}_{a,x},\{\sigma _{a|x}^{\prime }\}_{a,x})
= \sum_{a,x}p(x)D(\sigma _{a|x},\sigma _{a|x}^{\prime
})$.
First, it is obvious that the trace-distance measure of
steerability satisfies condition (i).
Before we prove that the trace-distance measure of steerability satisfies condition (ii), we prove the following Lemma.
\begin{lemma}.\label{lemma:noninc}
Let $\{\II_\omega\}_\omega$ be a collection of positive trace non-increasing maps, summing up to a trace non-increasing map $\II:=\sum_\omega \II_\omega$. Then,  for any Hermitian operators $T$ and $S$, we have
\begin{equation}
\sum_\omega \tr[| \II_\omega (T) - \II_\omega(S)| ]  \leq  \tr [|T-S|].
\end{equation}
\end{lemma}
{\it Proof.---}
The proof is a slight modification of the one by Ruskai~\cite{RUSKAI94}. Let us define $X:= T-S$. Since $X$ is Hermitian, by spectral decomposition, we can write $X=X^+ - X^-$, with $X^+,X^-\geq 0$. We then have
\begin{equation}
\begin{split}
\sum_\omega \tr[| \II_\omega (T) - \II_\omega(S)| ] = \sum_\omega \tr[ | \II_\omega (X) | ] \\
= \sum_\omega \tr[ | \II_\omega (X^+) - \II_\omega (X^-) | ] \\
\leq \sum_\omega \tr[ | \II_\omega (X^+)|] + \tr[| \II_\omega (X^-) | ]\\
=\sum_\omega \tr[  \II_\omega (X^+)] + \tr[ \II_\omega (X^-)  ]\\
=  \tr[ \sum_\omega \II_\omega (X^+ + X^-)  ]\\
\leq \tr[X^+ + X^-] =  \tr[| X^+ - X^-|] = \tr[|X|],
\end{split}
\end{equation}
where $\tr[| X|]$ and $\tr[ X]$ denote the trace norm and trace, respectively, and we used, in order, the triangle inequality, positivity, linearity, and trace non-increasing property. $\square$

\begin{lemma}.\label{lemma:wire}
The trace distance between two assemblages does not increase under deterministic wiring maps on Alice's side, under the restricted hypothesis $p(x|x',\omega)=p(x|x')$ of Eq.~\eqref{eq:wire_rest}.
\end{lemma}
{\it Proof.---} A wiring map $W_\omega$, depending on a parameter $\omega$, is a transformation of assemblages into assemblages given (component-wise) by
Eq. (4).
Note that given two assemblages $\{\sigma _{a|x}^1\}_{a,x},\{\sigma _{a|x}^2\}_{a,x}$, we can write
\begin{widetext}
\begin{equation}\label{eq:mon_wir}
\begin{split}
D_A({W}_\omega(\{\sigma_{a|x}^1\}_{a,x}), {W}_\omega (\{\sigma_{a|x}^2 \}_{a,x}) = D_A\left(\left\lbrace \sum_{a,x} p(x|x') p(a'|a,x,x',\omega)\sigma_{a|x}^1\right\rbrace_{a',x'}\ , \left\lbrace \sum_{a,x} p(x|x') p(a'|a,x,x',\omega)\sigma_{a|x}^2\right\rbrace_{a',x'}\right)\\
=  \sum_{a',x'} p(x') D\left(\sum_{a,x} p(x|x') p(a'|a,x,x',\omega) \sigma_{a|x}^1, \sum_{a,x} p(x|x') p(a'|a,x,x',\omega) \sigma_{a|x}^2\right) \\
\leq \sum_{a',x',a,x} p(x|x') p(a'|a,x,x',\omega) p(x') D(\sigma _{a|x}^1,\sigma _{a|x}^{2})
= \sum_{a,x,x'} p(x,x') D(\sigma _{a|x}^1,\sigma _{a|x}^{2})
\\ = \sum_{a,x} p(x) D(\sigma _{a|x}^1,\sigma _{a|x}^{2}) = D_{A}(\{\sigma _{a|x}^1\}_{a,x},\{\sigma _{a|x}^2\}_{a,x}),
\end{split}
\end{equation}
where the inequality holds since for $\lambda_i\geq 0$
(not necessarily summing up to one),
\begin{equation}\label{eq:almost_jc}
D(\sum_i \lambda_i \rho_i, \sum_i \lambda_i \rho'_i) = \frac{1}{2}\tr \left|\sum_i \lambda_i( \rho_i - \rho'_i)\right|
\leq \frac{1}{2}\sum_i \tr \left| \lambda_i( \rho_i - \rho'_i)\right| = \frac{1}{2}\sum_i \lambda_i \tr \left| \rho_i - \rho'_i \right|.
\end{equation}
This concludes the proof.$\square$

\begin{lemma}.\label{lemma:loc_op}
The quantifier $S_{\rm TD}$ does not increase, on average, by performing local operations on Bob's side defined by a collection of completely positive trace non-increasing maps $\{\II_\omega\}_\omega$, which sum up to a trace-preserving map $\II=\sum_\omega \II_\omega$.
\end{lemma}
{\it Proof.---} Let $\{\tilde{\rho}_{a|x}^{*\omega}\}_{a,x}$ be
the optimal unsteerable consistent assemblage giving the minimum
trace distance for $\II_\omega (\{ \sigma_{a|x} \}_{a,x}
)/P(\omega)$, and  $\{{\rho}_{a|x}^*\}_{a,x}$ the unsteerable
consistent assemblage giving the minimum trace-distance for $\{
\sigma_{a|x} \}_{a,x}$. We can then write
\begin{equation}
\begin{split}
\sum_\omega P(\omega) S_{\rm TD}\left(\frac{\II_\omega (\{ \sigma_{a|x} \}_{a,x} )}{P(\omega)} \right) =
\sum_\omega P(\omega) D_A \left(\frac{\II_\omega (\{ \sigma_{a|x} \}_{a,x} )}{P(\omega)}, \{ \tilde{\rho}_{a|x}^* \} \right) \hspace{5cm} \\
\leq  \sum_\omega P(\omega)D_A \left(\frac{\II_\omega (\{ \sigma_{a|x} \}_{a,x} )}{P(\omega)}, \frac{\II_\omega (\{ \rho_{a|x}^* \}_{a,x} )}{P(\omega)} \right) = \sum_\omega D_A \left({\II_\omega (\{ \sigma_{a|x} \}_{a,x} )}, \II_\omega (\{ \rho_{a|x}^* \}_{a,x} ) \right) \\
= \sum_{\omega,a,x} \tr \left[ \left| \II_\omega ( \sigma_{a|x}
)-\II_\omega (  \rho_{a|x}^*  ) \right| \right]\leq \sum_{a,x} \tr
\left[ \left|  \sigma_{a|x} - \rho_{a|x}^*   \right| \right]
\hspace{2cm}
\end{split}
\end{equation}
where we used for the first inequality the fact that $\tilde{\rho}_{a|x}^*$ is the minimum, linearity of the trace-distance for non-negative $P(\omega)$, and Lemma~\ref{lemma:noninc} for the last inequality. $\square$

\begin{theorem}.
The consistent trace-distance measure of steerability $S_{\rm TD}$ does not increase on average under restricted one-way LOCCs, namely
\begin{equation}
\sum_\omega P(\omega)\ S_{\rm TD}\left(\left\{ \frac{\sigma_{a'|x'}^\omega }{P(\omega)} \right\}_{a',x'} \right)\leq S_{\rm TD}(\{\sigma_{a|x}\}_{a,x}),
\end{equation}
\end{theorem}
\emph{Proof.---}The proof is simply given by an application of Lemmas~\ref{lemma:wire} and~\ref{lemma:loc_op}, namely

\begin{equation}
\begin{split}
\sum_\omega P(\omega)\ S_{\rm TD}\left(\left\{ \frac{\sigma_{a'|x'}^\omega }{P(\omega)} \right\}_{a',x'} \right) 
= \sum_\omega P(\omega)\ S_{\rm TD}\left(\left\{ \frac{W_\omega(\{K_\omega \sigma_{a|x} K_\omega^\dagger \}_{a,x})_{a'|x'} }{P(\omega)} \right\}_{a',x'} \right)\\
\leq \sum_\omega P(\omega)\ S_{\rm TD}\left(\left\{  \frac{K_\omega \sigma_{a|x} K_\omega^\dagger  }{P(\omega)} \right\}_{a,x} \right) \leq S_{\rm TD}(\{\sigma_{a|x}\}_{a,x}).
\end{split}
\end{equation}
$\square$

Finally, we prove convexity. Given the assemblage
$\{K_{a|x}\}_{a|x}$, obtained as a convex mixture $K_{a|x} = \mu
\sigma_{a|x} + (1-\mu ) \rho _{a|x}$, we have
\begin{equation}
\begin{split}
S_{\rm TD}(\{K_{a|x}\}_{a|x}) := \min_{\{K_{a|x}^{\text{US}}\}_{a,x}
\in\text{LHS}}D_{A}(\{{K_{a|x}}\}_{a,x},\{{K^{\text{US}}_{a|x}}\}_{a,x})\\
:=\min_{\{K_{a|x}^{\text{US}}\}_{a,x}\in\text{LHS}}D_{A}( \{\mu \sigma_{a|x} + (1-\mu ) \rho _{a|x}\}_{a,x},\{{K^{\text{US}}_{a|x}}\}_{a,x})\\
\leq D_{A}( \{\mu \sigma_{a|x} + (1-\mu ) \rho _{a|x}\}_{a,x},\{\mu \tilde{\sigma}^{\text{US}}_{a|x} + (1-\mu ) \tilde{\rho}^{\text{US}}_{a|x}\}_{a,x})\\
= \sum_{a,x} p(x)\frac{1}{2}\|\mu \sigma_{a|x} + (1-\mu ) \rho _{a|x}-\mu \tilde{\sigma}^{\text{US}}_{a|x} - (1-\mu ) \tilde{\rho}^{\text{US}}_{a|x}\|\\
= \sum_{a,x} p(x)\frac{1}{2}\|\mu \sigma_{a|x} -\mu \tilde{\sigma}^{\text{US}}_{a|x} + (1-\mu ) \rho _{a|x} - (1-\mu ) \tilde{\rho}^{\text{US}}_{a|x}\|\\
\leq \sum_{a,x}p(x)\Big\{ \frac{1}{2}\|\mu \sigma_{a|x} -\mu \tilde{\sigma}^{\text{US}}_{a|x}\|  +\frac{1}{2}\| (1-\mu ) \rho _{a|x} - (1-\mu ) \tilde{\rho}^{\text{US}}_{a|x}\|          \Big\}\\
= \mu S_{\rm{TD}}(\{\sigma_{a|x}\}_{a,x})+(1-\mu)S_{\rm{TD}}(\{\rho_{a|x}\}_{a,x}).
\end{split}
\end{equation}%
\end{widetext}
The first inequality is that the convex combination of the other two optimal LHS assemblages is not necessarily the  optimal assemblage for the convex combination assemblages (but it is still consistent in the sense of the total reduced state). The final inequality is due to the
property of the trace norm. This completes our proof of convexity.

\section{Semidefinite programming formulation of $S_{\rm CSR}^{\rm{R}}$ }\label{app:sdp}

The RNCSR,  $S_{\rm{CSR}}^{\rm{R}}$  defined by Eq.~(\ref{eq:srg}), can be
computed by the following SDP:
\begin{equation}\label{eq:SDPgamma}
\begin{split}
 \text{min: } & \sum_\lambda \tr(\sigma_\lambda) -1 \\
\text{subject to: }&
\sum_\lambda p(a|x,\lambda) \sigma_\lambda -\sigma_{a|x}\\
&\geq ( \sum_\lambda \tr(\sigma_\lambda) -1 ) \tr(\sigma_{a|x})\sigma_B ,\\
& \text{ for all } a,x;\\
  & \sum_\lambda \tr(\sigma_\lambda) \geq 1;\\
   & \sigma_\lambda \geq 0, \quad \forall \lambda;
\end{split}
\end{equation}
with $p(a|x,\lambda)$ taken as the deterministic strategies, i.e.,
$\lambda := (\lambda_x)_x$ and  $p(a|x,\lambda) :=
\delta_{a,\lambda_x}$.

In fact, it is sufficient to note that for all $t\geq S_{\rm
CSR}^{\rm{R}}$, there exists $\{\sigma'_\lambda\}_\lambda$ such
that
\begin{equation}
(1+t)\sum_\lambda p(a|x,\lambda) \sigma'_\lambda -\sigma_{a|x} = t
\tr(\sigma_{a|x})\sigma_B.
\end{equation}
Since $\sum_\lambda \tr(\sigma'_\lambda) = 1$, we can absorb the
factor ($1+t$) into the LHS assemblage, i.e., $\sigma_\lambda =
(1+t)\sigma'_\lambda$. Note that the first inequality in the
definition of the SDP, despite being a weaker condition than the
inequality, does not provide a lower value of $S_{\rm
CSR}^{\rm{R}}$. To prove this, let us just consider a feasible
solution $\{\sigma_\lambda\}$ of the SDP; we then have
\begin{eqnarray}
\sum_\lambda p(a|x,\lambda) \sigma_\lambda -\sigma_{a|x}-\Big[
\sum_\lambda \tr(\sigma_\lambda) -1 \Big] \tr(\sigma_{a|x})\sigma_B \nn\\ =:
\eta_{a|x} \geq 0.\quad
\end{eqnarray}
Then, by summing over $a$ and taking the trace of the left-hand
side of (C3), we obtain $\tr(\sum_a \eta_{a|x}) = 0$, for all $x$, which
implies $\eta_{a|x}=0$ for all $a,x$, since $\eta_{a|x} \geq 0$,
by our assumption.

\section{LHS surface of Werner states}\label{app:surf}

Here, we assume that one measurement is the $Z$ measurement and the others
are aligned in the $XY$ plane. Since three measurements are
orthogonal, the Werner states can be expanded in the Pauli bases, i.e.,
$\rho=\frac{1}{4}(\openone-\sum_{i=1}^{3}p\sigma_{i}\otimes\sigma_{i})$.
All projective measurements can also be expressed in the Pauli bases,
i.e., $E_{\pm|a}=\frac{1}{2}(\openone\pm\sum_{i=1}^{3}a_{i}\sigma_{i})$,
where $a$ can be seen as a vector in the Bloch sphere. Once Alice
performs projective measurements on her qubit, Bob's qubit
collapses into $\rho_{\pm}^{a}=\frac{1}{2}(\openone\pm\sum_{i}p
a_{i}\sigma_{i})$.

Now we use spherical coordinates to expand $a=(\sin \theta\cos
\phi,\sin\theta\sin\phi,\cos\theta)$. Alice can choose three
orthogonal $a_{i}$ given by
\begin{eqnarray}
 a_{1}&=&(\cos\phi,\sin\phi,0), \nonumber\\
 a_{2}&=&(\cos(\phi+\pi/2),\sin(\phi+\pi/2),0),\nonumber\\
 a_{3}&=&(0, 0, 1).
\label{N100}
\end{eqnarray}
The post-measurement states which Bob holds are
\begin{eqnarray}
\rho_{+}^{a_{1}}&=&(p\cos\phi,p\sin\phi,0),\nonumber\\
\rho_{-}^{a_{1}}&=&(-p\cos\phi,-p\sin\phi,0),\nonumber\\
\rho_{+}^{a_{2}}&=&(-p\sin\phi,p\cos\phi,0),\nonumber\\
\rho_{-}^{a_{2}}&=&(p\sin\phi,-p\cos\phi,0),\nonumber\\
\rho_{+}^{a_{3}}&=&(0,0,p),~~ \rho_{-}^{a_{3}}=(0,0,-p).
\label{N101}
\end{eqnarray}

Here we already use the Bloch-vector representation of a quantum
state. There are eight preexisted quantum states
$\sigma_{\lambda}$, which can be expressed as
\begin{eqnarray}
\sigma_{\lambda 1}=&p(\cos\phi-\sin\phi, \sin\phi+\cos\phi,1),\nonumber\\
\sigma_{\lambda 2}=&p(\cos\phi+\sin\phi, \sin\phi-\cos\phi,1),\nonumber\\
\sigma_{\lambda 3}=&p(-\cos\phi-\sin\phi, -\sin\phi+\cos\phi,1),\nonumber\\
\sigma_{\lambda 4}=&p(-\cos\phi+\sin\phi, -\sin\phi-\cos\phi,1),\nonumber\\
\sigma_{\lambda 5}=&p(\cos\phi-\sin\phi, \sin\phi+\cos\phi,-1),\nonumber\\
\sigma_{\lambda 6}=&p(\cos\phi+\sin\phi, \sin\phi-\cos\phi,-1),\nonumber\\
\sigma_{\lambda 7}=&p(-\cos\phi-\sin\phi, -\sin\phi+\cos\phi,-1),\nonumber\\
\sigma_{\lambda 8}=&p(-\cos\phi+\sin\phi, -\sin\phi-\cos\phi,-1)\nonumber.
\end{eqnarray}
It is obvious that the preexisted quantum states only exist when
$p\leq 1/\sqrt{3}$, because the radius of a pure-state Bloch
vector is equal to one. One can choose four specific preexisted
quantum states to mimic the post-measurement states with equal
probability of 1/4. Thus, the LHS states are
$\rho_{\pm}^{a_{i},\text{US}}$ and are given by Eqs.~(\ref{N101}),
with $p\leq 1/\sqrt{3}$ for $i=1,2,$ and $3$. As $p>1/\sqrt{3}$, the LHS states do not exist. We can
easily check that the states $\rho_{\pm}^{a_{i},\text{US}}$, given
by Eq.~(\ref{N101}) for $p\leq 1/\sqrt{3}$, are located at the
circle centered at $(0,0,0)$ and with radius $p$, because the
Werner states are highly symmetrical. Once we rotate the
measurement settings, the new measurements which correspond to the
original $XY$ plane are also located on a circle. Thus, the LHS
state of the Werner state is a sphere.


\begin{thebibliography}{74}%
\makeatletter
\providecommand \@ifxundefined [1]{%
 \@ifx{#1\undefined}
}%
\providecommand \@ifnum [1]{%
 \ifnum #1\expandafter \@firstoftwo
 \else \expandafter \@secondoftwo
 \fi
}%
\providecommand \@ifx [1]{%
 \ifx #1\expandafter \@firstoftwo
 \else \expandafter \@secondoftwo
 \fi
}%
\providecommand \natexlab [1]{#1}%
\providecommand \enquote  [1]{``#1''}%
\providecommand \bibnamefont  [1]{#1}%
\providecommand \bibfnamefont [1]{#1}%
\providecommand \citenamefont [1]{#1}%
\providecommand \href@noop [0]{\@secondoftwo}%
\providecommand \href [0]{\begingroup \@sanitize@url \@href}%
\providecommand \@href[1]{\@@startlink{#1}\@@href}%
\providecommand \@@href[1]{\endgroup#1\@@endlink}%
\providecommand \@sanitize@url [0]{\catcode `\\12\catcode `\$12\catcode
  `\&12\catcode `\#12\catcode `\^12\catcode `\_12\catcode `\%12\relax}%
\providecommand \@@startlink[1]{}%
\providecommand \@@endlink[0]{}%
\providecommand \url  [0]{\begingroup\@sanitize@url \@url }%
\providecommand \@url [1]{\endgroup\@href {#1}{\urlprefix }}%
\providecommand \urlprefix  [0]{URL }%
\providecommand \Eprint [0]{\href }%
\providecommand \doibase [0]{http://dx.doi.org/}%
\providecommand \selectlanguage [0]{\@gobble}%
\providecommand \bibinfo  [0]{\@secondoftwo}%
\providecommand \bibfield  [0]{\@secondoftwo}%
\providecommand \translation [1]{[#1]}%
\providecommand \BibitemOpen [0]{}%
\providecommand \bibitemStop [0]{}%
\providecommand \bibitemNoStop [0]{.\EOS\space}%
\providecommand \EOS [0]{\spacefactor3000\relax}%
\providecommand \BibitemShut  [1]{\csname bibitem#1\endcsname}%
\let\auto@bib@innerbib\@empty
\bibitem [{\citenamefont {Einstein}\ \emph {et~al.}(1935)\citenamefont
  {Einstein}, \citenamefont {Podolsky},\ and\ \citenamefont
  {Rosen}}]{Einstein35}%
  \BibitemOpen
  \bibfield  {author} {\bibinfo {author} {\bibfnamefont {A.}~\bibnamefont
  {Einstein}}, \bibinfo {author} {\bibfnamefont {B.}~\bibnamefont {Podolsky}},
  \ and\ \bibinfo {author} {\bibfnamefont {N.}~\bibnamefont {Rosen}},\
  }\bibfield  {title} {\enquote {\bibinfo {title} {Can quantum-mechanical
  description of physical reality be considered complete?}}\ }\href
  {http://link.aps.org/doi/10.1103/PhysRev.47.777} {\bibfield  {journal}
  {\bibinfo  {journal} {Phys. Rev.}\ }\textbf {\bibinfo {volume} {47}},\
  \bibinfo {pages} {777--780} (\bibinfo {year} {1935})}\BibitemShut {NoStop}%
\bibitem [{\citenamefont {Schr\"odinger}(1935)}]{Schrodinger35}%
  \BibitemOpen
  \bibfield  {author} {\bibinfo {author} {\bibfnamefont {E.}~\bibnamefont
  {Schr\"odinger}},\ }\bibfield  {title} {\enquote {\bibinfo {title}
  {Discussion of probability relations between separated systems},}\ }\href
  {http://journals.cambridge.org/article_S0305004100013554} {\bibfield
  {journal} {\bibinfo  {journal} {Proc. Cambridge Phil. Soc.}\ }\textbf
  {\bibinfo {volume} {31}},\ \bibinfo {pages} {555} (\bibinfo {year}
  {1935})}\BibitemShut {NoStop}%
\bibitem [{\citenamefont {Bell}(1964)}]{Bell64}%
  \BibitemOpen
  \bibfield  {author} {\bibinfo {author} {\bibfnamefont {J.~S.}\ \bibnamefont
  {Bell}},\ }\bibfield  {title} {\enquote {\bibinfo {title} {On the
  {E}instein-{P}odolsky-{R}osen paradox},}\ }\href
  {http://www.drchinese.com/David/Bell_Compact.pdf} {\bibfield  {journal}
  {\bibinfo  {journal} {Physics}\ }\textbf {\bibinfo {volume} {1}},\ \bibinfo
  {pages} {195--200} (\bibinfo {year} {1964})}\BibitemShut {NoStop}%
\bibitem [{\citenamefont {Wiseman}\ \emph {et~al.}(2007)\citenamefont
  {Wiseman}, \citenamefont {Jones},\ and\ \citenamefont {Doherty}}]{Wiseman07}%
  \BibitemOpen
  \bibfield  {author} {\bibinfo {author} {\bibfnamefont {H.~M.}\ \bibnamefont
  {Wiseman}}, \bibinfo {author} {\bibfnamefont {S.~J.}\ \bibnamefont {Jones}},
  \ and\ \bibinfo {author} {\bibfnamefont {A.~C.}\ \bibnamefont {Doherty}},\
  }\bibfield  {title} {\enquote {\bibinfo {title} {Steering, entanglement,
  nonlocality, and the {E}instein-{P}odolsky-{R}osen paradox},}\ }\href
  {http://link.aps.org/doi/10.1103/PhysRevLett.98.140402} {\bibfield  {journal}
  {\bibinfo  {journal} {Phys. Rev. Lett.}\ }\textbf {\bibinfo {volume} {98}},\
  \bibinfo {pages} {140402} (\bibinfo {year} {2007})}\BibitemShut {NoStop}%
\bibitem [{\citenamefont {Horodecki}\ \emph {et~al.}(2009)\citenamefont
  {Horodecki}, \citenamefont {Horodecki}, \citenamefont {Horodecki},\ and\
  \citenamefont {Horodecki}}]{Horodecki09RMP}%
  \BibitemOpen
  \bibfield  {author} {\bibinfo {author} {\bibfnamefont {R.}~\bibnamefont
  {Horodecki}}, \bibinfo {author} {\bibfnamefont {P.}~\bibnamefont
  {Horodecki}}, \bibinfo {author} {\bibfnamefont {M.}~\bibnamefont
  {Horodecki}}, \ and\ \bibinfo {author} {\bibfnamefont {K.}~\bibnamefont
  {Horodecki}},\ }\bibfield  {title} {\enquote {\bibinfo {title} {Quantum
  entanglement},}\ }\href {http://link.aps.org/doi/10.1103/RevModPhys.81.865}
  {\bibfield  {journal} {\bibinfo  {journal} {Rev. Mod. Phys.}\ }\textbf
  {\bibinfo {volume} {81}},\ \bibinfo {pages} {865--942} (\bibinfo {year}
  {2009})}\BibitemShut {NoStop}%
\bibitem [{\citenamefont {Brunner}\ \emph {et~al.}(2014)\citenamefont
  {Brunner}, \citenamefont {Cavalcanti}, \citenamefont {Pironio}, \citenamefont
  {Scarani},\ and\ \citenamefont {Wehner}}]{Brunner14RMP}%
  \BibitemOpen
  \bibfield  {author} {\bibinfo {author} {\bibfnamefont {N.}~\bibnamefont
  {Brunner}}, \bibinfo {author} {\bibfnamefont {D.}~\bibnamefont {Cavalcanti}},
  \bibinfo {author} {\bibfnamefont {S.}~\bibnamefont {Pironio}}, \bibinfo
  {author} {\bibfnamefont {V.}~\bibnamefont {Scarani}}, \ and\ \bibinfo
  {author} {\bibfnamefont {S.}~\bibnamefont {Wehner}},\ }\bibfield  {title}
  {\enquote {\bibinfo {title} {Bell nonlocality},}\ }\href
  {http://link.aps.org/doi/10.1103/RevModPhys.86.419} {\bibfield  {journal}
  {\bibinfo  {journal} {Rev. Mod. Phys.}\ }\textbf {\bibinfo {volume} {86}},\
  \bibinfo {pages} {419} (\bibinfo {year} {2014})}\BibitemShut {NoStop}%
\bibitem [{\citenamefont {Gallego}\ and\ \citenamefont
  {Aolita}(2015)}]{Gallego15}%
  \BibitemOpen
  \bibfield  {author} {\bibinfo {author} {\bibfnamefont {R.}~\bibnamefont
  {Gallego}}\ and\ \bibinfo {author} {\bibfnamefont {L.}~\bibnamefont
  {Aolita}},\ }\bibfield  {title} {\enquote {\bibinfo {title} {Resource theory
  of steering},}\ }\href {http://link.aps.org/doi/10.1103/PhysRevX.5.041008}
  {\bibfield  {journal} {\bibinfo  {journal} {Phys. Rev. X}\ }\textbf {\bibinfo
  {volume} {5}},\ \bibinfo {pages} {041008} (\bibinfo {year}
  {2015})}\BibitemShut {NoStop}%
\bibitem [{\citenamefont {Ou}\ \emph {et~al.}(1992)\citenamefont {Ou},
  \citenamefont {Pereira}, \citenamefont {Kimble},\ and\ \citenamefont
  {Peng}}]{Ou92}%
  \BibitemOpen
  \bibfield  {author} {\bibinfo {author} {\bibfnamefont {Z.~Y.}\ \bibnamefont
  {Ou}}, \bibinfo {author} {\bibfnamefont {S.~F.}\ \bibnamefont {Pereira}},
  \bibinfo {author} {\bibfnamefont {H.~J.}\ \bibnamefont {Kimble}}, \ and\
  \bibinfo {author} {\bibfnamefont {K.~C.}\ \bibnamefont {Peng}},\ }\bibfield
  {title} {\enquote {\bibinfo {title} {Realization of the
  {E}instein-{P}odolsky-{R}osen paradox for continuous variables},}\ }\href
  {http://dx.doi.org/10.1103/PhysRevLett.68.3663} {\bibfield  {journal}
  {\bibinfo  {journal} {Phys. Rev. Lett.}\ }\textbf {\bibinfo {volume} {68}},\
  \bibinfo {pages} {3663} (\bibinfo {year} {1992})}\BibitemShut {NoStop}%
\bibitem [{\citenamefont {Hald}\ \emph {et~al.}(1999)\citenamefont {Hald},
  \citenamefont {S\o{}rensen}, \citenamefont {Schori},\ and\ \citenamefont
  {Polzik}}]{Hald99}%
  \BibitemOpen
  \bibfield  {author} {\bibinfo {author} {\bibfnamefont {J.}~\bibnamefont
  {Hald}}, \bibinfo {author} {\bibfnamefont {J.~L.}\ \bibnamefont
  {S\o{}rensen}}, \bibinfo {author} {\bibfnamefont {C.}~\bibnamefont {Schori}},
  \ and\ \bibinfo {author} {\bibfnamefont {E.~S.}\ \bibnamefont {Polzik}},\
  }\bibfield  {title} {\enquote {\bibinfo {title} {Spin squeezed atoms: A
  macroscopic entangled ensemble created by light},}\ }\href
  {http://dx.doi.org/10.1103/PhysRevLett.83.1319} {\bibfield  {journal}
  {\bibinfo  {journal} {Phys. Rev. Lett.}\ }\textbf {\bibinfo {volume} {83}},\
  \bibinfo {pages} {1319} (\bibinfo {year} {1999})}\BibitemShut {NoStop}%
\bibitem [{\citenamefont {Bowen}\ \emph {et~al.}(2003)\citenamefont {Bowen},
  \citenamefont {Schnabel}, \citenamefont {Lam},\ and\ \citenamefont
  {Ralph}}]{Bowen03}%
  \BibitemOpen
  \bibfield  {author} {\bibinfo {author} {\bibfnamefont {W.~P.}\ \bibnamefont
  {Bowen}}, \bibinfo {author} {\bibfnamefont {R.}~\bibnamefont {Schnabel}},
  \bibinfo {author} {\bibfnamefont {P.~K.}\ \bibnamefont {Lam}}, \ and\
  \bibinfo {author} {\bibfnamefont {T.~C.}\ \bibnamefont {Ralph}},\ }\bibfield
  {title} {\enquote {\bibinfo {title} {Experimental investigation of criteria
  for continuous variable entanglement},}\ }\href
  {http://dx.doi.org/10.1103/PhysRevLett.90.043601} {\bibfield  {journal}
  {\bibinfo  {journal} {Phys. Rev. Lett.}\ }\textbf {\bibinfo {volume} {90}},\
  \bibinfo {pages} {043601} (\bibinfo {year} {2003})}\BibitemShut {NoStop}%
\bibitem [{\citenamefont {Howell}\ \emph {et~al.}(2004)\citenamefont {Howell},
  \citenamefont {Bennink}, \citenamefont {Bentley},\ and\ \citenamefont
  {Boyd}}]{Howell04}%
  \BibitemOpen
  \bibfield  {author} {\bibinfo {author} {\bibfnamefont {J.~C.}\ \bibnamefont
  {Howell}}, \bibinfo {author} {\bibfnamefont {R.~S.}\ \bibnamefont {Bennink}},
  \bibinfo {author} {\bibfnamefont {S.~J.}\ \bibnamefont {Bentley}}, \ and\
  \bibinfo {author} {\bibfnamefont {R.~W.}\ \bibnamefont {Boyd}},\ }\bibfield
  {title} {\enquote {\bibinfo {title} {Realization of the
  {E}instein-{P}odolsky-{R}osen paradox using momentum- and position-entangled
  photons from spontaneous parametric down conversion},}\ }\href
  {http://dx.doi.org/10.1103/PhysRevLett.92.210403} {\bibfield  {journal}
  {\bibinfo  {journal} {Phys. Rev. Lett.}\ }\textbf {\bibinfo {volume} {92}},\
  \bibinfo {pages} {210403} (\bibinfo {year} {2004})}\BibitemShut {NoStop}%
\bibitem [{\citenamefont {Saunders}\ \emph {et~al.}(2010)\citenamefont
  {Saunders}, \citenamefont {Jones}, \citenamefont {Wiseman},\ and\
  \citenamefont {Pryde}}]{Saunders10}%
  \BibitemOpen
  \bibfield  {author} {\bibinfo {author} {\bibfnamefont {D.~J.}\ \bibnamefont
  {Saunders}}, \bibinfo {author} {\bibfnamefont {S.~J.}\ \bibnamefont {Jones}},
  \bibinfo {author} {\bibfnamefont {H.~M.}\ \bibnamefont {Wiseman}}, \ and\
  \bibinfo {author} {\bibfnamefont {G.~J.}\ \bibnamefont {Pryde}},\ }\bibfield
  {title} {\enquote {\bibinfo {title} {Experimental {EPR}-steering using
  {B}ell-local states},}\ }\href {http://dx.doi.org/10.1038/nphys1766}
  {\bibfield  {journal} {\bibinfo  {journal} {Nat. Phys}\ }\textbf {\bibinfo
  {volume} {6}},\ \bibinfo {pages} {845--879} (\bibinfo {year}
  {2010})}\BibitemShut {NoStop}%
\bibitem [{\citenamefont {Wittmann}\ \emph {et~al.}(2012)\citenamefont
  {Wittmann}, \citenamefont {Ramelow}, \citenamefont {Steinlechner},
  \citenamefont {Langford}, \citenamefont {Brunner}, \citenamefont {Wiseman},
  \citenamefont {Ursin},\ and\ \citenamefont {Zeilinger}}]{Wittmann12}%
  \BibitemOpen
  \bibfield  {author} {\bibinfo {author} {\bibfnamefont {B.}~\bibnamefont
  {Wittmann}}, \bibinfo {author} {\bibfnamefont {S.}~\bibnamefont {Ramelow}},
  \bibinfo {author} {\bibfnamefont {F.}~\bibnamefont {Steinlechner}}, \bibinfo
  {author} {\bibfnamefont {N.~K.}\ \bibnamefont {Langford}}, \bibinfo {author}
  {\bibfnamefont {N.}~\bibnamefont {Brunner}}, \bibinfo {author} {\bibfnamefont
  {H.~M.}\ \bibnamefont {Wiseman}}, \bibinfo {author} {\bibfnamefont
  {R.}~\bibnamefont {Ursin}}, \ and\ \bibinfo {author} {\bibfnamefont
  {A.}~\bibnamefont {Zeilinger}},\ }\bibfield  {title} {\enquote {\bibinfo
  {title} {Loophole-free {E}instein-{P}odolsky-{R}osen experiment via quantum
  steering},}\ }\href {http://stacks.iop.org/1367-2630/14/i=5/a=053030}
  {\bibfield  {journal} {\bibinfo  {journal} {New J. Phys.}\ }\textbf {\bibinfo
  {volume} {14}},\ \bibinfo {pages} {053030} (\bibinfo {year}
  {2012})}\BibitemShut {NoStop}%
\bibitem [{\citenamefont {Bennet}\ \emph {et~al.}(2012)\citenamefont {Bennet},
  \citenamefont {Evans}, \citenamefont {Saunders}, \citenamefont {Branciard},
  \citenamefont {Cavalcanti}, \citenamefont {Wiseman},\ and\ \citenamefont
  {Pryde}}]{Bennet12}%
  \BibitemOpen
  \bibfield  {author} {\bibinfo {author} {\bibfnamefont {A.~J.}\ \bibnamefont
  {Bennet}}, \bibinfo {author} {\bibfnamefont {D.~A.}\ \bibnamefont {Evans}},
  \bibinfo {author} {\bibfnamefont {D.~J.}\ \bibnamefont {Saunders}}, \bibinfo
  {author} {\bibfnamefont {C.}~\bibnamefont {Branciard}}, \bibinfo {author}
  {\bibfnamefont {E.~G.}\ \bibnamefont {Cavalcanti}}, \bibinfo {author}
  {\bibfnamefont {H.~M.}\ \bibnamefont {Wiseman}}, \ and\ \bibinfo {author}
  {\bibfnamefont {G.~J.}\ \bibnamefont {Pryde}},\ }\bibfield  {title} {\enquote
  {\bibinfo {title} {Arbitrarily loss-tolerant {E}instein-{P}odolsky-{R}osen
  steering allowing a demonstration over 1~km of optical fiber with no
  detection loophole},}\ }\href {http://dx.doi.org/110.1103/PhysRevX.2.031003}
  {\bibfield  {journal} {\bibinfo  {journal} {Phys. Rev. X}\ }\textbf {\bibinfo
  {volume} {2}},\ \bibinfo {pages} {031003} (\bibinfo {year}
  {2012})}\BibitemShut {NoStop}%
\bibitem [{\citenamefont {H{\"a}ndchen}\ \emph {et~al.}(2012)\citenamefont
  {H{\"a}ndchen}, \citenamefont {Eberle}, \citenamefont {Steinlechner},
  \citenamefont {Samblowski}, \citenamefont {Franz}, \citenamefont {Werner},\
  and\ \citenamefont {Schnabel}}]{Handchen12}%
  \BibitemOpen
  \bibfield  {author} {\bibinfo {author} {\bibfnamefont {V.}~\bibnamefont
  {H{\"a}ndchen}}, \bibinfo {author} {\bibfnamefont {T.}~\bibnamefont
  {Eberle}}, \bibinfo {author} {\bibfnamefont {S.}~\bibnamefont
  {Steinlechner}}, \bibinfo {author} {\bibfnamefont {A.}~\bibnamefont
  {Samblowski}}, \bibinfo {author} {\bibfnamefont {T.}~\bibnamefont {Franz}},
  \bibinfo {author} {\bibfnamefont {R.~F.}\ \bibnamefont {Werner}}, \ and\
  \bibinfo {author} {\bibfnamefont {R.}~\bibnamefont {Schnabel}},\ }\bibfield
  {title} {\enquote {\bibinfo {title} {Observation of one-way
  {E}instein-{P}odolsky-{R}osen steering},}\ }\href
  {http://dx.doi.org/11010.1038/nphoton.2012.202} {\bibfield  {journal}
  {\bibinfo  {journal} {Nat. Photon.}\ }\textbf {\bibinfo {volume} {6}},\
  \bibinfo {pages} {596--599} (\bibinfo {year} {2012})}\BibitemShut {NoStop}%
\bibitem [{\citenamefont {Smith}\ \emph {et~al.}(2012)\citenamefont {Smith},
  \citenamefont {Gillett}, \citenamefont {de~Almeida}, \citenamefont
  {Branciard}, \citenamefont {Fedrizzi}, \citenamefont {Weinhold},
  \citenamefont {Lita}, \citenamefont {Calkins}, \citenamefont {Gerrits},
  \citenamefont {Wiseman}, \citenamefont {Nam},\ and\ \citenamefont
  {White}}]{Smith12}%
  \BibitemOpen
  \bibfield  {author} {\bibinfo {author} {\bibfnamefont {D.~H.}\ \bibnamefont
  {Smith}}, \bibinfo {author} {\bibfnamefont {G.}~\bibnamefont {Gillett}},
  \bibinfo {author} {\bibfnamefont {M.~P.}\ \bibnamefont {de~Almeida}},
  \bibinfo {author} {\bibfnamefont {C.}~\bibnamefont {Branciard}}, \bibinfo
  {author} {\bibfnamefont {A.}~\bibnamefont {Fedrizzi}}, \bibinfo {author}
  {\bibfnamefont {T.~J.}\ \bibnamefont {Weinhold}}, \bibinfo {author}
  {\bibfnamefont {A.}~\bibnamefont {Lita}}, \bibinfo {author} {\bibfnamefont
  {B.}~\bibnamefont {Calkins}}, \bibinfo {author} {\bibfnamefont
  {T.}~\bibnamefont {Gerrits}}, \bibinfo {author} {\bibfnamefont {H.~M.}\
  \bibnamefont {Wiseman}}, \bibinfo {author} {\bibfnamefont {S.~W.}\
  \bibnamefont {Nam}}, \ and\ \bibinfo {author} {\bibfnamefont {A.~G.}\
  \bibnamefont {White}},\ }\bibfield  {title} {\enquote {\bibinfo {title}
  {Conclusive quantum steering with superconducting transition-edge sensors},}\
  }\href {http://dx.doi.org/10.1038/ncomms1628} {\bibfield  {journal} {\bibinfo
   {journal} {Nat. Comm.}\ }\textbf {\bibinfo {volume} {3}},\ \bibinfo {pages}
  {845} (\bibinfo {year} {2012})}\BibitemShut {NoStop}%
\bibitem [{\citenamefont {Steinlechner}\ \emph {et~al.}(2013)\citenamefont
  {Steinlechner}, \citenamefont {Bauchrowitz}, \citenamefont {Eberle},\ and\
  \citenamefont {Schnabel}}]{Steinlechner13}%
  \BibitemOpen
  \bibfield  {author} {\bibinfo {author} {\bibfnamefont {S.}~\bibnamefont
  {Steinlechner}}, \bibinfo {author} {\bibfnamefont {J.}~\bibnamefont
  {Bauchrowitz}}, \bibinfo {author} {\bibfnamefont {T.}~\bibnamefont {Eberle}},
  \ and\ \bibinfo {author} {\bibfnamefont {R.}~\bibnamefont {Schnabel}},\
  }\bibfield  {title} {\enquote {\bibinfo {title} {Strong
  {E}instein-{P}odolsky-{R}osen steering with unconditional entangled
  states},}\ }\href {http://dx.doi.org/110.1103/PhysRevA.87.022104} {\bibfield
  {journal} {\bibinfo  {journal} {Phys. Rev. A}\ }\textbf {\bibinfo {volume}
  {87}},\ \bibinfo {pages} {022104} (\bibinfo {year} {2013})}\BibitemShut
  {NoStop}%
\bibitem [{\citenamefont {Schneeloch}\ \emph {et~al.}(2013)\citenamefont
  {Schneeloch}, \citenamefont {Dixon}, \citenamefont {Howland}, \citenamefont
  {Broadbent},\ and\ \citenamefont {Howell}}]{Schneeloch13}%
  \BibitemOpen
  \bibfield  {author} {\bibinfo {author} {\bibfnamefont {J.}~\bibnamefont
  {Schneeloch}}, \bibinfo {author} {\bibfnamefont {P.~B.}\ \bibnamefont
  {Dixon}}, \bibinfo {author} {\bibfnamefont {G.~A.}\ \bibnamefont {Howland}},
  \bibinfo {author} {\bibfnamefont {C.~J.}\ \bibnamefont {Broadbent}}, \ and\
  \bibinfo {author} {\bibfnamefont {J.~C.}\ \bibnamefont {Howell}},\ }\bibfield
   {title} {\enquote {\bibinfo {title} {Violation of continuous-variable
  {E}instein-{P}odolsky-{R}osen steering with discrete measurements},}\ }\href
  {http://dx.doi.org/110.1103/PhysRevLett.110.130407} {\bibfield  {journal}
  {\bibinfo  {journal} {Phys. Rev. Lett.}\ }\textbf {\bibinfo {volume} {110}},\
  \bibinfo {pages} {130407} (\bibinfo {year} {2013})}\BibitemShut {NoStop}%
\bibitem [{\citenamefont {Su}\ \emph {et~al.}(2013)\citenamefont {Su},
  \citenamefont {Chen}, \citenamefont {Wu}, \citenamefont {Deng},\ and\
  \citenamefont {Oh}}]{Su13}%
  \BibitemOpen
  \bibfield  {author} {\bibinfo {author} {\bibfnamefont {H.-Y.}\ \bibnamefont
  {Su}}, \bibinfo {author} {\bibfnamefont {J.-L.}\ \bibnamefont {Chen}},
  \bibinfo {author} {\bibfnamefont {C.}~\bibnamefont {Wu}}, \bibinfo {author}
  {\bibfnamefont {D.-L.}\ \bibnamefont {Deng}}, \ and\ \bibinfo {author}
  {\bibfnamefont {C.~H.}\ \bibnamefont {Oh}},\ }\bibfield  {title} {\enquote
  {\bibinfo {title} {Detecting {E}instein-{P}odolsky-{R}osen steering for
  continuous variable wavefunctions},}\ }\href
  {http://dx.doi.org/110.1142/S0219749913500196} {\bibfield  {journal}
  {\bibinfo  {journal} {I. J. Quant. Infor.}\ }\textbf {\bibinfo {volume}
  {11}},\ \bibinfo {pages} {1350019} (\bibinfo {year} {2013})}\BibitemShut
  {NoStop}%
\bibitem [{\citenamefont {Sun}\ \emph {et~al.}(2016)\citenamefont {Sun},
  \citenamefont {Ye}, \citenamefont {Xu}, \citenamefont {Xu}, \citenamefont
  {Tang}, \citenamefont {Wu}, \citenamefont {Chen}, \citenamefont {Li},\ and\
  \citenamefont {Guo}}]{Sun16}%
  \BibitemOpen
  \bibfield  {author} {\bibinfo {author} {\bibfnamefont {K.}~\bibnamefont
  {Sun}}, \bibinfo {author} {\bibfnamefont {X.-J.}\ \bibnamefont {Ye}},
  \bibinfo {author} {\bibfnamefont {J.-S.}\ \bibnamefont {Xu}}, \bibinfo
  {author} {\bibfnamefont {X.-Y.}\ \bibnamefont {Xu}}, \bibinfo {author}
  {\bibfnamefont {J.-S.}\ \bibnamefont {Tang}}, \bibinfo {author}
  {\bibfnamefont {Y.-C.}\ \bibnamefont {Wu}}, \bibinfo {author} {\bibfnamefont
  {J.-L.}\ \bibnamefont {Chen}}, \bibinfo {author} {\bibfnamefont {C.-F.}\
  \bibnamefont {Li}}, \ and\ \bibinfo {author} {\bibfnamefont {G.-C.}\
  \bibnamefont {Guo}},\ }\bibfield  {title} {\enquote {\bibinfo {title}
  {Experimental quantification of asymmetric {E}instein-{P}odolsky-{R}osen
  steering},}\ }\href {http://link.aps.org/doi/10.1103/PhysRevLett.116.160404}
  {\bibfield  {journal} {\bibinfo  {journal} {Phys. Rev. Lett.}\ }\textbf
  {\bibinfo {volume} {116}},\ \bibinfo {pages} {160404} (\bibinfo {year}
  {2016})}\BibitemShut {NoStop}%
\bibitem [{\citenamefont {Cavalcanti}\ \emph {et~al.}(2009)\citenamefont
  {Cavalcanti}, \citenamefont {Jones}, \citenamefont {Wiseman},\ and\
  \citenamefont {Reid}}]{Cavalcanti09}%
  \BibitemOpen
  \bibfield  {author} {\bibinfo {author} {\bibfnamefont {E.~G.}\ \bibnamefont
  {Cavalcanti}}, \bibinfo {author} {\bibfnamefont {S.~J.}\ \bibnamefont
  {Jones}}, \bibinfo {author} {\bibfnamefont {H.~M.}\ \bibnamefont {Wiseman}},
  \ and\ \bibinfo {author} {\bibfnamefont {M.~D.}\ \bibnamefont {Reid}},\
  }\bibfield  {title} {\enquote {\bibinfo {title} {Experimental criteria for
  steering and the {E}instein-{P}odolsky-{R}osen paradox},}\ }\href
  {http://link.aps.org/doi/10.1103/PhysRevA.80.032112} {\bibfield  {journal}
  {\bibinfo  {journal} {Phys. Rev. A}\ }\textbf {\bibinfo {volume} {80}},\
  \bibinfo {pages} {032112} (\bibinfo {year} {2009})}\BibitemShut {NoStop}%
\bibitem [{\citenamefont {Reid}(1989)}]{Reid89}%
  \BibitemOpen
  \bibfield  {author} {\bibinfo {author} {\bibfnamefont {M.~D.}\ \bibnamefont
  {Reid}},\ }\bibfield  {title} {\enquote {\bibinfo {title} {Demonstration of
  the {E}instein-{P}odolsky-{R}osen paradox using nondegenerate parametric
  amplification},}\ }\href {http://link.aps.org/doi/10.1103/PhysRevA.40.913}
  {\bibfield  {journal} {\bibinfo  {journal} {Phys. Rev. A}\ }\textbf {\bibinfo
  {volume} {40}},\ \bibinfo {pages} {913--923} (\bibinfo {year}
  {1989})}\BibitemShut {NoStop}%
\bibitem [{\citenamefont {Pusey}(2013)}]{Pusey13}%
  \BibitemOpen
  \bibfield  {author} {\bibinfo {author} {\bibfnamefont {M.~F.}\ \bibnamefont
  {Pusey}},\ }\bibfield  {title} {\enquote {\bibinfo {title} {Negativity and
  steering: A stronger {P}eres conjecture},}\ }\href
  {http://link.aps.org/doi/10.1103/PhysRevA.88.032313} {\bibfield  {journal}
  {\bibinfo  {journal} {Phys. Rev. A}\ }\textbf {\bibinfo {volume} {88}},\
  \bibinfo {pages} {032313} (\bibinfo {year} {2013})}\BibitemShut {NoStop}%
\bibitem [{\citenamefont {Walborn}\ \emph {et~al.}(2011)\citenamefont
  {Walborn}, \citenamefont {Salles}, \citenamefont {Gomes}, \citenamefont
  {Toscano},\ and\ \citenamefont {Souto~Ribeiro}}]{Walborn11}%
  \BibitemOpen
  \bibfield  {author} {\bibinfo {author} {\bibfnamefont {S.~P.}\ \bibnamefont
  {Walborn}}, \bibinfo {author} {\bibfnamefont {A.}~\bibnamefont {Salles}},
  \bibinfo {author} {\bibfnamefont {R.~M.}\ \bibnamefont {Gomes}}, \bibinfo
  {author} {\bibfnamefont {F.}~\bibnamefont {Toscano}}, \ and\ \bibinfo
  {author} {\bibfnamefont {P.~H.}\ \bibnamefont {Souto~Ribeiro}},\ }\bibfield
  {title} {\enquote {\bibinfo {title} {Revealing hidden
  {E}instein-{P}odolsky-{R}osen nonlocality},}\ }\href
  {http://dx.doi.org/110.1103/PhysRevLett.106.130402} {\bibfield  {journal}
  {\bibinfo  {journal} {Phys. Rev. Lett.}\ }\textbf {\bibinfo {volume} {106}},\
  \bibinfo {pages} {130402} (\bibinfo {year} {2011})}\BibitemShut {NoStop}%
\bibitem [{\citenamefont {Kogias}\ \emph {et~al.}(2015)\citenamefont {Kogias},
  \citenamefont {Lee}, \citenamefont {Ragy},\ and\ \citenamefont
  {Adesso}}]{Kogias15}%
  \BibitemOpen
  \bibfield  {author} {\bibinfo {author} {\bibfnamefont {I.}~\bibnamefont
  {Kogias}}, \bibinfo {author} {\bibfnamefont {A.~R.}\ \bibnamefont {Lee}},
  \bibinfo {author} {\bibfnamefont {S.}~\bibnamefont {Ragy}}, \ and\ \bibinfo
  {author} {\bibfnamefont {G.}~\bibnamefont {Adesso}},\ }\bibfield  {title}
  {\enquote {\bibinfo {title} {Quantification of {G}aussian quantum
  steering},}\ }\href {http://link.aps.org/doi/10.1103/PhysRevLett.114.060403}
  {\bibfield  {journal} {\bibinfo  {journal} {Phys. Rev. Lett.}\ }\textbf
  {\bibinfo {volume} {114}},\ \bibinfo {pages} {060403} (\bibinfo {year}
  {2015})}\BibitemShut {NoStop}%
\bibitem [{\citenamefont {Costa}\ and\ \citenamefont {Angelo}(2016)}]{Costa16}%
  \BibitemOpen
  \bibfield  {author} {\bibinfo {author} {\bibfnamefont {A.~C.~S.}\
  \bibnamefont {Costa}}\ and\ \bibinfo {author} {\bibfnamefont {R.~M.}\
  \bibnamefont {Angelo}},\ }\bibfield  {title} {\enquote {\bibinfo {title}
  {Quantification of {E}instein-{P}odolski-{R}osen steering for two-qubit
  states},}\ }\href {http://link.aps.org/doi/10.1103/PhysRevA.93.020103}
  {\bibfield  {journal} {\bibinfo  {journal} {Phys. Rev. A}\ }\textbf {\bibinfo
  {volume} {93}},\ \bibinfo {pages} {020103} (\bibinfo {year}
  {2016})}\BibitemShut {NoStop}%
\bibitem [{\citenamefont {He}\ and\ \citenamefont {Reid}(2013)}]{He13}%
  \BibitemOpen
  \bibfield  {author} {\bibinfo {author} {\bibfnamefont {Q.~Y.}\ \bibnamefont
  {He}}\ and\ \bibinfo {author} {\bibfnamefont {M.~D.}\ \bibnamefont {Reid}},\
  }\bibfield  {title} {\enquote {\bibinfo {title} {Genuine multipartite
  {E}instein-{P}odolsky-{R}osen steering},}\ }\href
  {http://link.aps.org/doi/10.1103/PhysRevLett.111.250403} {\bibfield
  {journal} {\bibinfo  {journal} {Phys. Rev. Lett.}\ }\textbf {\bibinfo
  {volume} {111}},\ \bibinfo {pages} {250403} (\bibinfo {year}
  {2013})}\BibitemShut {NoStop}%
\bibitem [{\citenamefont {Li}\ \emph {et~al.}(2015{\natexlab{a}})\citenamefont
  {Li}, \citenamefont {Chen}, \citenamefont {Chen}, \citenamefont {Zhang},
  \citenamefont {Chen},\ and\ \citenamefont {Pan}}]{Li_Che_Ming15}%
  \BibitemOpen
  \bibfield  {author} {\bibinfo {author} {\bibfnamefont {C.-M}\ \bibnamefont
  {Li}}, \bibinfo {author} {\bibfnamefont {K.}~\bibnamefont {Chen}}, \bibinfo
  {author} {\bibfnamefont {Y.-N.}\ \bibnamefont {Chen}}, \bibinfo {author}
  {\bibfnamefont {Q.}~\bibnamefont {Zhang}}, \bibinfo {author} {\bibfnamefont
  {Y.-A}\ \bibnamefont {Chen}}, \ and\ \bibinfo {author} {\bibfnamefont
  {J.-W.}\ \bibnamefont {Pan}},\ }\bibfield  {title} {\enquote {\bibinfo
  {title} {Genuine high-order {E}instein-{P}odolsky-{R}osen steering},}\ }\href
  {http://link.aps.org/doi/10.1103/PhysRevLett.115.010402} {\bibfield
  {journal} {\bibinfo  {journal} {Phys. Rev. Lett.}\ }\textbf {\bibinfo
  {volume} {115}},\ \bibinfo {pages} {010402} (\bibinfo {year}
  {2015}{\natexlab{a}})}\BibitemShut {NoStop}%
\bibitem [{\citenamefont {Xiang}\ \emph {et~al.}(2017)\citenamefont {Xiang},
  \citenamefont {Kogias}, \citenamefont {Adesso},\ and\ \citenamefont
  {He}}]{Xiang17}%
  \BibitemOpen
  \bibfield  {author} {\bibinfo {author} {\bibfnamefont {Y.}~\bibnamefont
  {Xiang}}, \bibinfo {author} {\bibfnamefont {I.}~\bibnamefont {Kogias}},
  \bibinfo {author} {\bibfnamefont {G.}~\bibnamefont {Adesso}}, \ and\ \bibinfo
  {author} {\bibfnamefont {Q.}~\bibnamefont {He}},\ }\bibfield  {title}
  {\enquote {\bibinfo {title} {Multipartite {G}aussian steering: Monogamy
  constraints and quantum cryptography applications},}\ }\href
  {http://link.aps.org/doi/10.1103/PhysRevA.95.010101} {\bibfield  {journal}
  {\bibinfo  {journal} {Phys. Rev. A}\ }\textbf {\bibinfo {volume} {95}},\
  \bibinfo {pages} {010101} (\bibinfo {year} {2017})}\BibitemShut {NoStop}%
\bibitem [{\citenamefont {Milne}\ \emph {et~al.}(2014)\citenamefont {Milne},
  \citenamefont {Jevtic}, \citenamefont {Jennings}, \citenamefont {Wiseman},\
  and\ \citenamefont {Rudolph}}]{Milne14}%
  \BibitemOpen
  \bibfield  {author} {\bibinfo {author} {\bibfnamefont {A.}~\bibnamefont
  {Milne}}, \bibinfo {author} {\bibfnamefont {S.}~\bibnamefont {Jevtic}},
  \bibinfo {author} {\bibfnamefont {D.}~\bibnamefont {Jennings}}, \bibinfo
  {author} {\bibfnamefont {H.}~\bibnamefont {Wiseman}}, \ and\ \bibinfo
  {author} {\bibfnamefont {T.}~\bibnamefont {Rudolph}},\ }\bibfield  {title}
  {\enquote {\bibinfo {title} {Quantum steering ellipsoids, extremal physical
  states and monogamy},}\ }\href
  {http://stacks.iop.org/1367-2630/16/i=8/a=083017} {\bibfield  {journal}
  {\bibinfo  {journal} {New J. Phys.}\ }\textbf {\bibinfo {volume} {16}},\
  \bibinfo {pages} {083017} (\bibinfo {year} {2014})}\BibitemShut {NoStop}%
\bibitem [{\citenamefont {Cheng}\ \emph {et~al.}(2016)\citenamefont {Cheng},
  \citenamefont {Milne}, \citenamefont {Hall},\ and\ \citenamefont
  {Wiseman}}]{Cheng16}%
  \BibitemOpen
  \bibfield  {author} {\bibinfo {author} {\bibfnamefont {S.}~\bibnamefont
  {Cheng}}, \bibinfo {author} {\bibfnamefont {A.}~\bibnamefont {Milne}},
  \bibinfo {author} {\bibfnamefont {M.~J.~W.}\ \bibnamefont {Hall}}, \ and\
  \bibinfo {author} {\bibfnamefont {H.~M.}\ \bibnamefont {Wiseman}},\
  }\bibfield  {title} {\enquote {\bibinfo {title} {Volume monogamy of quantum
  steering ellipsoids for multiqubit systems},}\ }\href
  {http://link.aps.org/doi/10.1103/PhysRevA.94.042105} {\bibfield  {journal}
  {\bibinfo  {journal} {Phys. Rev. A}\ }\textbf {\bibinfo {volume} {94}},\
  \bibinfo {pages} {042105} (\bibinfo {year} {2016})}\BibitemShut {NoStop}%
\bibitem [{\citenamefont {Cavalcanti}\ and\ \citenamefont
  {Skrzypczyk}(2016)}]{Cavalcanti16}%
  \BibitemOpen
  \bibfield  {author} {\bibinfo {author} {\bibfnamefont {D.}~\bibnamefont
  {Cavalcanti}}\ and\ \bibinfo {author} {\bibfnamefont {P.}~\bibnamefont
  {Skrzypczyk}},\ }\bibfield  {title} {\enquote {\bibinfo {title} {Quantitative
  relations between measurement incompatibility, quantum steering, and
  nonlocality},}\ }\href {http://link.aps.org/doi/10.1103/PhysRevA.93.052112}
  {\bibfield  {journal} {\bibinfo  {journal} {Phys. Rev. A}\ }\textbf {\bibinfo
  {volume} {93}},\ \bibinfo {pages} {052112} (\bibinfo {year}
  {2016})}\BibitemShut {NoStop}%
\bibitem [{\citenamefont {Uola}\ \emph {et~al.}(2014)\citenamefont {Uola},
  \citenamefont {Moroder},\ and\ \citenamefont {G\"uhne}}]{Uola14}%
  \BibitemOpen
  \bibfield  {author} {\bibinfo {author} {\bibfnamefont {R.}~\bibnamefont
  {Uola}}, \bibinfo {author} {\bibfnamefont {T.}~\bibnamefont {Moroder}}, \
  and\ \bibinfo {author} {\bibfnamefont {O.}~\bibnamefont {G\"uhne}},\
  }\bibfield  {title} {\enquote {\bibinfo {title} {Joint measurability of
  generalized measurements implies classicality},}\ }\href
  {http://link.aps.org/doi/10.1103/PhysRevLett.113.160403} {\bibfield
  {journal} {\bibinfo  {journal} {Phys. Rev. Lett.}\ }\textbf {\bibinfo
  {volume} {113}},\ \bibinfo {pages} {160403} (\bibinfo {year}
  {2014})}\BibitemShut {NoStop}%
\bibitem [{\citenamefont {Quintino}\ \emph {et~al.}(2014)\citenamefont
  {Quintino}, \citenamefont {V\'ertesi},\ and\ \citenamefont
  {Brunner}}]{Quintino14}%
  \BibitemOpen
  \bibfield  {author} {\bibinfo {author} {\bibfnamefont {M.~T.}\
  \bibnamefont {Quintino}}, \bibinfo {author} {\bibfnamefont {T.}~\bibnamefont
  {V\'ertesi}}, \ and\ \bibinfo {author} {\bibfnamefont {N.}~\bibnamefont
  {Brunner}},\ }\bibfield  {title} {\enquote {\bibinfo {title} {Joint
  measurability, {E}instein-{P}odolsky-{R}osen steering, and {B}ell
  nonlocality},}\ }\href
  {http://link.aps.org/doi/10.1103/PhysRevLett.113.160402} {\bibfield
  {journal} {\bibinfo  {journal} {Phys. Rev. Lett.}\ }\textbf {\bibinfo
  {volume} {113}},\ \bibinfo {pages} {160402} (\bibinfo {year}
  {2014})}\BibitemShut {NoStop}%
\bibitem [{\citenamefont {Chen}\ \emph
  {et~al.}(2016{\natexlab{a}})\citenamefont {Chen}, \citenamefont {Budroni},
  \citenamefont {Liang},\ and\ \citenamefont {Chen}}]{Shin-Liang16c}%
  \BibitemOpen
  \bibfield  {author} {\bibinfo {author} {\bibfnamefont {S.-L.}\ \bibnamefont
  {Chen}}, \bibinfo {author} {\bibfnamefont {C.}~\bibnamefont {Budroni}},
  \bibinfo {author} {\bibfnamefont {Y.-C.}\ \bibnamefont {Liang}}, \ and\
  \bibinfo {author} {\bibfnamefont {Y.-N.}\ \bibnamefont {Chen}},\ }\bibfield
  {title} {\enquote {\bibinfo {title} {Natural framework for device-independent
  quantification of quantum steerability, measurement incompatibility, and
  self-testing},}\ }\href
  {http://link.aps.org/doi/10.1103/PhysRevLett.116.240401} {\bibfield
  {journal} {\bibinfo  {journal} {Phys. Rev. Lett.}\ }\textbf {\bibinfo
  {volume} {116}},\ \bibinfo {pages} {240401} (\bibinfo {year}
  {2016}{\natexlab{a}})}\BibitemShut {NoStop}%
\bibitem [{\citenamefont {Uola}\ \emph {et~al.}(2015)\citenamefont {Uola},
  \citenamefont {Budroni}, \citenamefont {G\"uhne},\ and\ \citenamefont
  {Pellonp\"a\"a}}]{Uola15}%
  \BibitemOpen
  \bibfield  {author} {\bibinfo {author} {\bibfnamefont {R.}~\bibnamefont
  {Uola}}, \bibinfo {author} {\bibfnamefont {C.}~\bibnamefont {Budroni}},
  \bibinfo {author} {\bibfnamefont {O.}~\bibnamefont {G\"uhne}}, \ and\
  \bibinfo {author} {\bibfnamefont {J.~P.}~\bibnamefont {Pellonp\"a\"a}},\
  }\bibfield  {title} {\enquote {\bibinfo {title} {One-to-one mapping between
  steering and joint measurability problems},}\ }\href
  {http://link.aps.org/doi/10.1103/PhysRevLett.115.230402} {\bibfield
  {journal} {\bibinfo  {journal} {Phys. Rev. Lett.}\ }\textbf {\bibinfo
  {volume} {115}},\ \bibinfo {pages} {230402} (\bibinfo {year}
  {2015})}\BibitemShut {NoStop}%
\bibitem [{\citenamefont {Wollmann}\ \emph {et~al.}(2016)\citenamefont
  {Wollmann}, \citenamefont {Walk}, \citenamefont {Bennet}, \citenamefont
  {Wiseman},\ and\ \citenamefont {Pryde}}]{Wollmann16}%
  \BibitemOpen
  \bibfield  {author} {\bibinfo {author} {\bibfnamefont {S.}~\bibnamefont
  {Wollmann}}, \bibinfo {author} {\bibfnamefont {N.}~\bibnamefont {Walk}},
  \bibinfo {author} {\bibfnamefont {A.~J.}\ \bibnamefont {Bennet}}, \bibinfo
  {author} {\bibfnamefont {H.~M.}\ \bibnamefont {Wiseman}}, \ and\ \bibinfo
  {author} {\bibfnamefont {G.~J.}\ \bibnamefont {Pryde}},\ }\bibfield  {title}
  {\enquote {\bibinfo {title} {Observation of genuine one-way
  {E}instein-{P}odolsky-{R}osen steering},}\ }\href
  {http://link.aps.org/doi/10.1103/PhysRevLett.116.160403} {\bibfield
  {journal} {\bibinfo  {journal} {Phys. Rev. Lett.}\ }\textbf {\bibinfo
  {volume} {116}},\ \bibinfo {pages} {160403} (\bibinfo {year}
  {2016})}\BibitemShut {NoStop}%
\bibitem [{\citenamefont {Skrzypczyk}\ \emph {et~al.}(2014)\citenamefont
  {Skrzypczyk}, \citenamefont {Navascu\'es},\ and\ \citenamefont
  {Cavalcanti}}]{Skrzypczyk14}%
  \BibitemOpen
  \bibfield  {author} {\bibinfo {author} {\bibfnamefont {P.}~\bibnamefont
  {Skrzypczyk}}, \bibinfo {author} {\bibfnamefont {M.}~\bibnamefont
  {Navascu\'es}}, \ and\ \bibinfo {author} {\bibfnamefont {D.}~\bibnamefont
  {Cavalcanti}},\ }\bibfield  {title} {\enquote {\bibinfo {title} {Quantifying
  {E}instein-{P}odolsky-{R}osen steering},}\ }\href
  {http://link.aps.org/doi/10.1103/PhysRevLett.112.180404} {\bibfield
  {journal} {\bibinfo  {journal} {Phys. Rev. Lett.}\ }\textbf {\bibinfo
  {volume} {112}},\ \bibinfo {pages} {180404} (\bibinfo {year}
  {2014})}\BibitemShut {NoStop}%
\bibitem [{\citenamefont {Cavalcanti}\ and\ \citenamefont
  {Skrzypczyk}(2017)}]{SDPreview17}%
  \BibitemOpen
  \bibfield  {author} {\bibinfo {author} {\bibfnamefont {D.}~\bibnamefont
  {Cavalcanti}}\ and\ \bibinfo {author} {\bibfnamefont {P.}~\bibnamefont
  {Skrzypczyk}},\ }\bibfield  {title} {\enquote {\bibinfo {title} {Quantum
  steering: a review with focus on semidefinite programming},}\ }\href
  {http://stacks.iop.org/0034-4885/80/i=2/a=024001} {\bibfield  {journal}
  {\bibinfo  {journal} {Rep. Prog. Phys.}\ }\textbf {\bibinfo {volume} {80}},\
  \bibinfo {pages} {024001} (\bibinfo {year} {2017})}\BibitemShut {NoStop}%
\bibitem [{\citenamefont {Branciard}\ \emph {et~al.}(2012)\citenamefont
  {Branciard}, \citenamefont {Cavalcanti}, \citenamefont {Walborn},
  \citenamefont {Scarani},\ and\ \citenamefont {Wiseman}}]{Branciard12}%
  \BibitemOpen
  \bibfield  {author} {\bibinfo {author} {\bibfnamefont {C.}~\bibnamefont
  {Branciard}}, \bibinfo {author} {\bibfnamefont {E.~G.}\ \bibnamefont
  {Cavalcanti}}, \bibinfo {author} {\bibfnamefont {S.~P.}\ \bibnamefont
  {Walborn}}, \bibinfo {author} {\bibfnamefont {V.}~\bibnamefont {Scarani}}, \
  and\ \bibinfo {author} {\bibfnamefont {H.~M.}\ \bibnamefont {Wiseman}},\
  }\bibfield  {title} {\enquote {\bibinfo {title} {One-sided device-independent
  quantum key distribution: Security, feasibility, and the connection with
  steering},}\ }\href {http://dx.doi.org/110.1103/PhysRevA.85.010301}
  {\bibfield  {journal} {\bibinfo  {journal} {Phys. Rev. A}\ }\textbf {\bibinfo
  {volume} {85}},\ \bibinfo {pages} {010301} (\bibinfo {year}
  {2012})}\BibitemShut {NoStop}%
\bibitem [{\citenamefont {Tatham}\ \emph {et~al.}(2012)\citenamefont {Tatham},
  \citenamefont {Mi\ifmmode~\check{s}\else \v{s}\fi{}ta}, \citenamefont
  {Adesso},\ and\ \citenamefont {Korolkova}}]{Tatham12}%
  \BibitemOpen
  \bibfield  {author} {\bibinfo {author} {\bibfnamefont {R.}~\bibnamefont
  {Tatham}}, \bibinfo {author} {\bibfnamefont {L.}~\bibnamefont
  {Mi\ifmmode~\check{s}\else \v{s}\fi{}ta}}, \bibinfo {author} {\bibfnamefont
  {G.}~\bibnamefont {Adesso}}, \ and\ \bibinfo {author} {\bibfnamefont
  {N.}~\bibnamefont {Korolkova}},\ }\bibfield  {title} {\enquote {\bibinfo
  {title} {Nonclassical correlations in continuous-variable non-{G}aussian
  {W}erner states},}\ }\href
  {http://link.aps.org/doi/10.1103/PhysRevA.85.022326} {\bibfield  {journal}
  {\bibinfo  {journal} {Phys. Rev. A}\ }\textbf {\bibinfo {volume} {85}},\
  \bibinfo {pages} {022326} (\bibinfo {year} {2012})}\BibitemShut {NoStop}%
\bibitem [{\citenamefont {He}\ \emph {et~al.}(2015)\citenamefont {He},
  \citenamefont {Rosales-Z\'arate}, \citenamefont {Adesso},\ and\ \citenamefont
  {Reid}}]{Qiongyi15}%
  \BibitemOpen
  \bibfield  {author} {\bibinfo {author} {\bibfnamefont {Q.}~\bibnamefont
  {He}}, \bibinfo {author} {\bibfnamefont {L.}~\bibnamefont
  {Rosales-Z\'arate}}, \bibinfo {author} {\bibfnamefont {G.}~\bibnamefont
  {Adesso}}, \ and\ \bibinfo {author} {\bibfnamefont {M.~D.}\ \bibnamefont
  {Reid}},\ }\bibfield  {title} {\enquote {\bibinfo {title} {Secure continuous
  variable teleportation and {E}instein-{P}odolsky-{R}osen steering},}\ }\href
  {http://link.aps.org/doi/10.1103/PhysRevLett.115.180502} {\bibfield
  {journal} {\bibinfo  {journal} {Phys. Rev. Lett.}\ }\textbf {\bibinfo
  {volume} {115}},\ \bibinfo {pages} {180502} (\bibinfo {year}
  {2015})}\BibitemShut {NoStop}%
\bibitem [{\citenamefont {Wang}\ \emph {et~al.}(2015)\citenamefont {Wang},
  \citenamefont {Xiang}, \citenamefont {He},\ and\ \citenamefont
  {Gong}}]{Wang15}%
  \BibitemOpen
  \bibfield  {author} {\bibinfo {author} {\bibfnamefont {M.}~\bibnamefont
  {Wang}}, \bibinfo {author} {\bibfnamefont {Y.}~\bibnamefont {Xiang}},
  \bibinfo {author} {\bibfnamefont {Q.}~\bibnamefont {He}}, \ and\ \bibinfo
  {author} {\bibfnamefont {Q.}~\bibnamefont {Gong}},\ }\bibfield  {title}
  {\enquote {\bibinfo {title} {Asymmetric quantum network based on multipartite
  {E}instein-{P}odolsky-{R}osen steering},}\ }\href
  {http://josab.osa.org/abstract.cfm?URI=josab-32-4-A20} {\bibfield  {journal}
  {\bibinfo  {journal} {J. Opt. Soc. Am. B}\ }\textbf {\bibinfo {volume}
  {32}},\ \bibinfo {pages} {A20--A26} (\bibinfo {year} {2015})}\BibitemShut
  {NoStop}%
\bibitem [{\citenamefont {Chen}\ \emph {et~al.}(2014)\citenamefont {Chen},
  \citenamefont {Li}, \citenamefont {Lambert}, \citenamefont {Chen},
  \citenamefont {Ota}, \citenamefont {Chen},\ and\ \citenamefont
  {Nori}}]{Yueh-Nan14}%
  \BibitemOpen
  \bibfield  {author} {\bibinfo {author} {\bibfnamefont {Y.-N.}\ \bibnamefont
  {Chen}}, \bibinfo {author} {\bibfnamefont {C.-M.}\ \bibnamefont {Li}},
  \bibinfo {author} {\bibfnamefont {N.}~\bibnamefont {Lambert}}, \bibinfo
  {author} {\bibfnamefont {S.-L.}\ \bibnamefont {Chen}}, \bibinfo {author}
  {\bibfnamefont {Y.}~\bibnamefont {Ota}}, \bibinfo {author} {\bibfnamefont
  {G.-Y.}\ \bibnamefont {Chen}}, \ and\ \bibinfo {author} {\bibfnamefont
  {F.}~\bibnamefont {Nori}},\ }\bibfield  {title} {\enquote {\bibinfo {title}
  {Temporal steering inequality},}\ }\href
  {http://link.aps.org/doi/10.1103/PhysRevA.89.032112} {\bibfield  {journal}
  {\bibinfo  {journal} {Phys. Rev. A}\ }\textbf {\bibinfo {volume} {89}},\
  \bibinfo {pages} {032112} (\bibinfo {year} {2014})}\BibitemShut {NoStop}%
\bibitem [{\citenamefont {Chen}\ \emph
  {et~al.}(2016{\natexlab{b}})\citenamefont {Chen}, \citenamefont {Lambert},
  \citenamefont {Li}, \citenamefont {Miranowicz}, \citenamefont {Chen},\ and\
  \citenamefont {Nori}}]{Shin-Liang16}%
  \BibitemOpen
  \bibfield  {author} {\bibinfo {author} {\bibfnamefont {S.-L.}\ \bibnamefont
  {Chen}}, \bibinfo {author} {\bibfnamefont {N.}~\bibnamefont {Lambert}},
  \bibinfo {author} {\bibfnamefont {C.-M.}\ \bibnamefont {Li}}, \bibinfo
  {author} {\bibfnamefont {A.}~\bibnamefont {Miranowicz}}, \bibinfo {author}
  {\bibfnamefont {Y.-N.}\ \bibnamefont {Chen}}, \ and\ \bibinfo {author}
  {\bibfnamefont {F.}~\bibnamefont {Nori}},\ }\bibfield  {title} {\enquote
  {\bibinfo {title} {Quantifying non-{M}arkovianity with temporal steering},}\
  }\href {http://link.aps.org/doi/10.1103/PhysRevLett.116.020503} {\bibfield
  {journal} {\bibinfo  {journal} {Phys. Rev. Lett.}\ }\textbf {\bibinfo
  {volume} {116}},\ \bibinfo {pages} {020503} (\bibinfo {year}
  {2016}{\natexlab{b}})}\BibitemShut {NoStop}%
\bibitem [{\citenamefont {Bartkiewicz}\ \emph {et~al.}(2016)\citenamefont
  {Bartkiewicz}, \citenamefont {\ifmmode~\check{C}\else \v{C}\fi{}ernoch},
  \citenamefont {Lemr}, \citenamefont {Miranowicz},\ and\ \citenamefont
  {Nori}}]{Bartkiewicz16a}%
  \BibitemOpen
  \bibfield  {author} {\bibinfo {author} {\bibfnamefont {K.}~\bibnamefont
  {Bartkiewicz}}, \bibinfo {author} {\bibfnamefont {A.}~\bibnamefont
  {\ifmmode~\check{C}\else \v{C}\fi{}ernoch}}, \bibinfo {author} {\bibfnamefont
  {K.}~\bibnamefont {Lemr}}, \bibinfo {author} {\bibfnamefont {A.}~\bibnamefont
  {Miranowicz}}, \ and\ \bibinfo {author} {\bibfnamefont {F.}~\bibnamefont
  {Nori}},\ }\bibfield  {title} {\enquote {\bibinfo {title} {Temporal steering
  and security of quantum key distribution with mutually unbiased bases against
  individual attacks},}\ }\href
  {http://link.aps.org/doi/10.1103/PhysRevA.93.062345} {\bibfield  {journal}
  {\bibinfo  {journal} {Phys. Rev. A}\ }\textbf {\bibinfo {volume} {93}},\
  \bibinfo {pages} {062345} (\bibinfo {year} {2016})}\BibitemShut {NoStop}%
\bibitem [{\citenamefont {{Bartkiewicz}}\ \emph {et~al.}(2016)\citenamefont
  {{Bartkiewicz}}, \citenamefont {{{\v C}ernoch}}, \citenamefont {{Lemr}},
  \citenamefont {{Miranowicz}},\ and\ \citenamefont {{Nori}}}]{Bartkiewicz16b}%
  \BibitemOpen
  \bibfield  {author} {\bibinfo {author} {\bibfnamefont {K.}~\bibnamefont
  {{Bartkiewicz}}}, \bibinfo {author} {\bibfnamefont {A.}~\bibnamefont {{{\v
  C}ernoch}}}, \bibinfo {author} {\bibfnamefont {K.}~\bibnamefont {{Lemr}}},
  \bibinfo {author} {\bibfnamefont {A.}~\bibnamefont {{Miranowicz}}}, \ and\
  \bibinfo {author} {\bibfnamefont {F.}~\bibnamefont {{Nori}}},\ }\bibfield
  {title} {\enquote {\bibinfo {title} {Experimental temporal quantum
  steering},}\ }\href {https://doi.org/10.1038/srep38076} {\bibfield  {journal}
  {\bibinfo  {journal} {Sc. Rep.}\ }\textbf {\bibinfo {volume} {6}},\ \bibinfo
  {pages} {38076} (\bibinfo {year} {2016})}\BibitemShut {NoStop}%
\bibitem [{\citenamefont {Ku}\ \emph {et~al.}(2016)\citenamefont {Ku},
  \citenamefont {Chen}, \citenamefont {Chen}, \citenamefont {Lambert},
  \citenamefont {Chen},\ and\ \citenamefont {Nori}}]{Ku16}%
  \BibitemOpen
  \bibfield  {author} {\bibinfo {author} {\bibfnamefont {H.-Y.}\ \bibnamefont
  {Ku}}, \bibinfo {author} {\bibfnamefont {S.-L.}\ \bibnamefont {Chen}},
  \bibinfo {author} {\bibfnamefont {H.-B.}\ \bibnamefont {Chen}}, \bibinfo
  {author} {\bibfnamefont {N.}~\bibnamefont {Lambert}}, \bibinfo {author}
  {\bibfnamefont {Y.-N.}\ \bibnamefont {Chen}}, \ and\ \bibinfo {author}
  {\bibfnamefont {F.}~\bibnamefont {Nori}},\ }\bibfield  {title} {\enquote
  {\bibinfo {title} {Temporal steering in four dimensions with applications to
  coupled qubits and magnetoreception},}\ }\href
  {http://link.aps.org/doi/10.1103/PhysRevA.94.062126} {\bibfield  {journal}
  {\bibinfo  {journal} {Phys. Rev. A}\ }\textbf {\bibinfo {volume} {94}},\
  \bibinfo {pages} {062126} (\bibinfo {year} {2016})}\BibitemShut {NoStop}%
\bibitem [{\citenamefont {Li}\ \emph {et~al.}(2015{\natexlab{b}})\citenamefont
  {Li}, \citenamefont {Chen}, \citenamefont {Lambert}, \citenamefont {Chiu},\
  and\ \citenamefont {Nori}}]{Che-Ming15}%
  \BibitemOpen
  \bibfield  {author} {\bibinfo {author} {\bibfnamefont {C.-M.}\ \bibnamefont
  {Li}}, \bibinfo {author} {\bibfnamefont {Y.-N.}\ \bibnamefont {Chen}},
  \bibinfo {author} {\bibfnamefont {N.}~\bibnamefont {Lambert}}, \bibinfo
  {author} {\bibfnamefont {C.-Y.}~\bibnamefont {Chiu}}, \ and\ \bibinfo {author}
  {\bibfnamefont {F.}~\bibnamefont {Nori}},\ }\bibfield  {title} {\enquote
  {\bibinfo {title} {Certifying single-system steering for quantum-information
  processing},}\ }\href {http://link.aps.org/doi/10.1103/PhysRevA.92.062310}
  {\bibfield  {journal} {\bibinfo  {journal} {Phys. Rev. A}\ }\textbf {\bibinfo
  {volume} {92}},\ \bibinfo {pages} {062310} (\bibinfo {year}
  {2015}{\natexlab{b}})}\BibitemShut {NoStop}%
\bibitem [{\citenamefont {Chen}\ \emph {et~al.}(2017)\citenamefont {Chen},
  \citenamefont {Lambert}, \citenamefont {Li}, \citenamefont {Chen},
  \citenamefont {Chen}, \citenamefont {Miranowicz},\ and\ \citenamefont
  {Nori}}]{Chen2017}%
  \BibitemOpen
  \bibfield  {author} {\bibinfo {author} {\bibfnamefont {S.-L.}\ \bibnamefont
  {Chen}}, \bibinfo {author} {\bibfnamefont {N.}~\bibnamefont {Lambert}},
  \bibinfo {author} {\bibfnamefont {C.-M.}\ \bibnamefont {Li}}, \bibinfo
  {author} {\bibfnamefont {G.-Yin.}\ \bibnamefont {Chen}}, \bibinfo {author}
  {\bibfnamefont {Y.-N.}\ \bibnamefont {Chen}}, \bibinfo {author}
  {\bibfnamefont {A.}~\bibnamefont {Miranowicz}}, \ and\ \bibinfo {author}
  {\bibfnamefont {F.}~\bibnamefont {Nori}},\ }\bibfield  {title} {\enquote
  {\bibinfo {title} {Spatio-temporal steering for testing nonclassical
  correlations in quantum networks},}\ }\href
  {https://doi.org/10.1038/s41598-017-03789-4} {\bibfield  {journal} {\bibinfo
  {journal} {Sci. Rep.}\ }\textbf {\bibinfo {volume} {7}},\ \bibinfo {pages} {3728} (\bibinfo {year}
  {2017})}\BibitemShut {NoStop}%
\bibitem [{\citenamefont {Piani}\ and\ \citenamefont
  {Watrous}(2015)}]{Piani15}%
  \BibitemOpen
  \bibfield  {author} {\bibinfo {author} {\bibfnamefont {M.}~\bibnamefont
  {Piani}}\ and\ \bibinfo {author} {\bibfnamefont {J.}~\bibnamefont
  {Watrous}},\ }\bibfield  {title} {\enquote {\bibinfo {title} {Necessary and
  sufficient quantum information characterization of
  {E}instein-{P}odolsky-{R}osen steering},}\ }\href
  {http://link.aps.org/doi/10.1103/PhysRevLett.114.060404} {\bibfield
  {journal} {\bibinfo  {journal} {Phys. Rev. Lett.}\ }\textbf {\bibinfo
  {volume} {114}},\ \bibinfo {pages} {060404} (\bibinfo {year}
  {2015})}\BibitemShut {NoStop}%
\bibitem [{\citenamefont {Hsieh}\ \emph {et~al.}(2016)\citenamefont {Hsieh},
  \citenamefont {Liang},\ and\ \citenamefont {Lee}}]{Hsieh16}%
  \BibitemOpen
  \bibfield  {author} {\bibinfo {author} {\bibfnamefont {C.-Y.}\ \bibnamefont
  {Hsieh}}, \bibinfo {author} {\bibfnamefont {Y.-C.}\ \bibnamefont {Liang}}, \
  and\ \bibinfo {author} {\bibfnamefont {R.-K.}\ \bibnamefont {Lee}},\
  }\bibfield  {title} {\enquote {\bibinfo {title} {Quantum steerability:
  Characterization, quantification, superactivation, and unbounded
  amplification},}\ }\href {http://link.aps.org/doi/10.1103/PhysRevA.94.062120}
  {\bibfield  {journal} {\bibinfo  {journal} {Phys. Rev. A}\ }\textbf {\bibinfo
  {volume} {94}},\ \bibinfo {pages} {062120} (\bibinfo {year}
  {2016})}\BibitemShut {NoStop}%
\bibitem [{\citenamefont {Das}\ \emph {et~al.}(2017)\citenamefont {Das},
  \citenamefont {Datta}, \citenamefont {Jebaratnam},\ and\ \citenamefont
  {Majumdar}}]{Das17}%
  \BibitemOpen
  \bibfield  {author} {\bibinfo {author} {\bibfnamefont {D.}~\bibnamefont
  {Das}}, \bibinfo {author} {\bibfnamefont {S.}~\bibnamefont {Datta}}, \bibinfo
  {author} {\bibfnamefont {C.}~\bibnamefont {Jebaratnam}}, \ and\ \bibinfo
  {author} {\bibfnamefont {A.~S.}\ \bibnamefont {Majumdar}},\ }\bibfield
  {title} {\enquote {\bibinfo {title} {Cost of Einstein-Podolsky-Rosen steering in the context of extremal
boxes},}\ }\href {https://link.aps.org/doi/10.1103/PhysRevA.97.022110}
  {\bibfield  {journal} {\bibinfo  {journal} {Phys. Rev. A}\ }\textbf {\bibinfo
  {volume} {97}},\ \bibinfo {pages} {022110} (\bibinfo {year}
  {2018})}\BibitemShut {NoStop}%
\bibitem [{\citenamefont {Kaur}\ \emph {et~al.}(2017)\citenamefont {Kaur},
  \citenamefont {Wang},\ and\ \citenamefont {Wilde}}]{Eneet17a}%
  \BibitemOpen
  \bibfield  {author} {\bibinfo {author} {\bibfnamefont {E.}~\bibnamefont
  {Kaur}}, \bibinfo {author} {\bibfnamefont {X.}~\bibnamefont {Wang}}, \ and\
  \bibinfo {author} {\bibfnamefont {M.~M.}\ \bibnamefont {Wilde}},\ }\bibfield
  {title} {\enquote {\bibinfo {title} {Conditional mutual information and
  quantum steering},}\ }\href {\doibase 10.1103/PhysRevA.96.022332} {\bibfield
  {journal} {\bibinfo  {journal} {Phys. Rev. A}\ }\textbf {\bibinfo {volume}
  {96}},\ \bibinfo {pages} {022332} (\bibinfo {year} {2017})}\BibitemShut
  {NoStop}%
\bibitem [{\citenamefont {Kaur}\ and\ \citenamefont {Wilde}(2017)}]{Eneet17b}%
  \BibitemOpen
  \bibfield  {author} {\bibinfo {author} {\bibfnamefont {E.}~\bibnamefont
  {Kaur}}\ and\ \bibinfo {author} {\bibfnamefont {M.~M.}\ \bibnamefont
  {Wilde}},\ }\bibfield  {title} {\enquote {\bibinfo {title} {Relative entropy
  of steering: on its definition and properties},}\ }\href
  {http://stacks.iop.org/1751-8121/50/i=46/a=465301} {\bibfield  {journal}
  {\bibinfo  {journal} {J. Phys. A}\ }\textbf {\bibinfo {volume} {50}},\
  \bibinfo {pages} {465301} (\bibinfo {year} {2017})}\BibitemShut {NoStop}%
\bibitem [{\citenamefont {Jevtic}\ \emph {et~al.}(2014)\citenamefont {Jevtic},
  \citenamefont {Pusey}, \citenamefont {Jennings},\ and\ \citenamefont
  {Rudolph}}]{Jevtic14}%
  \BibitemOpen
  \bibfield  {author} {\bibinfo {author} {\bibfnamefont {S.}~\bibnamefont
  {Jevtic}}, \bibinfo {author} {\bibfnamefont {M.}~\bibnamefont {Pusey}},
  \bibinfo {author} {\bibfnamefont {D.}~\bibnamefont {Jennings}}, \ and\
  \bibinfo {author} {\bibfnamefont {T.}~\bibnamefont {Rudolph}},\ }\bibfield
  {title} {\enquote {\bibinfo {title} {Quantum steering ellipsoids},}\ }\href
  {http://link.aps.org/doi/10.1103/PhysRevLett.113.020402} {\bibfield
  {journal} {\bibinfo  {journal} {Phys. Rev. Lett.}\ }\textbf {\bibinfo
  {volume} {113}},\ \bibinfo {pages} {020402} (\bibinfo {year}
  {2014})}\BibitemShut {NoStop}%
\bibitem [{\citenamefont {Jevtic}\ \emph {et~al.}(2015)\citenamefont {Jevtic},
  \citenamefont {Hall}, \citenamefont {Anderson}, \citenamefont {Zwierz},\ and\
  \citenamefont {Wiseman}}]{Jevtic15}%
  \BibitemOpen
  \bibfield  {author} {\bibinfo {author} {\bibfnamefont {S.}~\bibnamefont
  {Jevtic}}, \bibinfo {author} {\bibfnamefont {M.~J.~W.}\ \bibnamefont {Hall}},
  \bibinfo {author} {\bibfnamefont {M.~R.}\ \bibnamefont {Anderson}}, \bibinfo
  {author} {\bibfnamefont {M.}~\bibnamefont {Zwierz}}, \ and\ \bibinfo {author}
  {\bibfnamefont {H.~M.}\ \bibnamefont {Wiseman}},\ }\bibfield  {title}
  {\enquote {\bibinfo {title} {Einstein--{P}odolsky--{R}osen steering and the
  steering ellipsoid},}\ }\href
  {http://josab.osa.org/abstract.cfm?URI=josab-32-4-A40} {\bibfield  {journal}
  {\bibinfo  {journal} {J. Opt. Soc. Am. B}\ }\textbf {\bibinfo {volume}
  {32}},\ \bibinfo {pages} {A40--A49} (\bibinfo {year} {2015})}\BibitemShut
  {NoStop}%
\bibitem [{\citenamefont {Nguyen}\ and\ \citenamefont
  {Vu}(2016{\natexlab{a}})}]{Nguyen16a}%
  \BibitemOpen
  \bibfield  {author} {\bibinfo {author} {\bibfnamefont {H.~C.}\ \bibnamefont
  {Nguyen}}\ and\ \bibinfo {author} {\bibfnamefont {T.}~\bibnamefont {Vu}},\
  }\bibfield  {title} {\enquote {\bibinfo {title} {Necessary and sufficient
  condition for steerability of two-qubit states by the geometry of steering
  outcomes},}\ }\href {http://stacks.iop.org/0295-5075/115/i=1/a=10003}
  {\bibfield  {journal} {\bibinfo  {journal} {Europhysics Letters}\ }\textbf
  {\bibinfo {volume} {115}},\ \bibinfo {pages} {10003} (\bibinfo {year}
  {2016}{\natexlab{a}})}\BibitemShut {NoStop}%
\bibitem [{\citenamefont {Nguyen}\ and\ \citenamefont
  {Vu}(2016{\natexlab{b}})}]{Nguyen16b}%
  \BibitemOpen
  \bibfield  {author} {\bibinfo {author} {\bibfnamefont {H.~C.}\ \bibnamefont
  {Nguyen}}\ and\ \bibinfo {author} {\bibfnamefont {T.}~\bibnamefont {Vu}},\
  }\bibfield  {title} {\enquote {\bibinfo {title} {Nonseparability and
  steerability of two-qubit states from the geometry of steering outcomes},}\
  }\href {http://link.aps.org/doi/10.1103/PhysRevA.94.012114} {\bibfield
  {journal} {\bibinfo  {journal} {Phys. Rev. A}\ }\textbf {\bibinfo {volume}
  {94}},\ \bibinfo {pages} {012114} (\bibinfo {year}
  {2016}{\natexlab{b}})}\BibitemShut {NoStop}%
\bibitem [{\citenamefont {McCloskey}\ \emph {et~al.}(2017)\citenamefont
  {McCloskey}, \citenamefont {Ferraro},\ and\ \citenamefont
  {Paternostro}}]{McCloskey16}%
  \BibitemOpen
  \bibfield  {author} {\bibinfo {author} {\bibfnamefont {R.}~\bibnamefont
  {McCloskey}}, \bibinfo {author} {\bibfnamefont {A.}~\bibnamefont {Ferraro}},
  \ and\ \bibinfo {author} {\bibfnamefont {M.}~\bibnamefont {Paternostro}},\
  }\bibfield  {title} {\enquote {\bibinfo {title} {Einstein-{P}odolsky-{R}osen
  steering and quantum steering ellipsoids: Optimal two-qubit states and
  projective measurements},}\ }\href
  {https://doi.org/10.1103/physreva.95.012320} {\bibfield  {journal} {\bibinfo
  {journal} {Phys. Rev. A}\ }\textbf {\bibinfo {volume} {95}},\ \bibinfo {pages} {012320} (\bibinfo {year}
  {2017})}\BibitemShut {NoStop}%
\bibitem [{\citenamefont {Bowles}\ \emph {et~al.}(2016)\citenamefont {Bowles},
  \citenamefont {Hirsch}, \citenamefont {Quintino},\ and\ \citenamefont
  {Brunner}}]{Bowles16}%
  \BibitemOpen
  \bibfield  {author} {\bibinfo {author} {\bibfnamefont {J.}~\bibnamefont
  {Bowles}}, \bibinfo {author} {\bibfnamefont {F.}~\bibnamefont {Hirsch}},
  \bibinfo {author} {\bibfnamefont {M.~T.}\ \bibnamefont {Quintino}}, \ and\
  \bibinfo {author} {\bibfnamefont {N.}~\bibnamefont {Brunner}},\ }\bibfield
  {title} {\enquote {\bibinfo {title} {Sufficient criterion for guaranteeing
  that a two-qubit state is unsteerable},}\ }\href
  {http://link.aps.org/doi/10.1103/PhysRevA.93.022121} {\bibfield  {journal}
  {\bibinfo  {journal} {Phys. Rev. A}\ }\textbf {\bibinfo {volume} {93}},\
  \bibinfo {pages} {022121} (\bibinfo {year} {2016})}\BibitemShut {NoStop}%
\bibitem [{\citenamefont {Hillery}(1987)}]{Hillery87}%
  \BibitemOpen
  \bibfield  {author} {\bibinfo {author} {\bibfnamefont {M.}~\bibnamefont
  {Hillery}},\ }\bibfield  {title} {\enquote {\bibinfo {title} {Nonclassical
  distance in quantum optics},}\ }\href
  {http://dx.doi.org/10.1103/PhysRevA.35.725} {\bibfield  {journal} {\bibinfo
  {journal} {Phys. Rev. A}\ }\textbf {\bibinfo {volume} {35}},\ \bibinfo
  {pages} {725} (\bibinfo {year} {1987})}\BibitemShut {NoStop}%
\bibitem [{\citenamefont {Miranowicz}\ \emph {et~al.}(2015)\citenamefont
  {Miranowicz}, \citenamefont {Bartkiewicz}, \citenamefont {Pathak},
  \citenamefont {Pe{\v{r}}ina}, \citenamefont {Chen},\ and\ \citenamefont
  {Nori}}]{Adam15}%
  \BibitemOpen
  \bibfield  {author} {\bibinfo {author} {\bibfnamefont {A.}~\bibnamefont
  {Miranowicz}}, \bibinfo {author} {\bibfnamefont {K.}~\bibnamefont
  {Bartkiewicz}}, \bibinfo {author} {\bibfnamefont {A.}~\bibnamefont {Pathak}},
  \bibinfo {author} {\bibfnamefont {J.}~\bibnamefont {Pe{\v{r}}ina}}, \bibinfo
  {author} {\bibfnamefont {Y.-N.}\ \bibnamefont {Chen}}, \ and\ \bibinfo
  {author} {\bibfnamefont {F.}~\bibnamefont {Nori}},\ }\bibfield  {title}
  {\enquote {\bibinfo {title} {Statistical mixtures of states can be more
  quantum than their superpositions: {C}omparison of nonclassicality measures
  for single-qubit states},}\ }\href
  {http://dx.doi.org/10.1103/PhysRevA.91.042309} {\bibfield  {journal}
  {\bibinfo  {journal} {Phys. Rev. A}\ }\textbf {\bibinfo {volume} {91}},\ \bibinfo {pages} {042309}
  (\bibinfo {year} {2015})}\BibitemShut {NoStop}%
\bibitem [{\citenamefont {Vedral}\ \emph {et~al.}(1997)\citenamefont {Vedral},
  \citenamefont {Plenio}, \citenamefont {Rippin},\ and\ \citenamefont
  {Knight}}]{Vedral97}%
  \BibitemOpen
  \bibfield  {author} {\bibinfo {author} {\bibfnamefont {V.}~\bibnamefont
  {Vedral}}, \bibinfo {author} {\bibfnamefont {M.~B.}\ \bibnamefont {Plenio}},
  \bibinfo {author} {\bibfnamefont {M.~A.}\ \bibnamefont {Rippin}}, \ and\
  \bibinfo {author} {\bibfnamefont {P.~L.}\ \bibnamefont {Knight}},\ }\bibfield
   {title} {\enquote {\bibinfo {title} {Quantifying entanglement},}\ }\href
  {http://dx.doi.org/10.1103/PhysRevLett.78.2275} {\bibfield  {journal}
  {\bibinfo  {journal} {Phys. Rev. Lett.}\ }\textbf {\bibinfo {volume} {78}},\
  \bibinfo {pages} {2275} (\bibinfo {year} {1997})}\BibitemShut {NoStop}%
\bibitem [{\citenamefont {Rana}\ \emph {et~al.}(2016)\citenamefont {Rana},
  \citenamefont {Parashar},\ and\ \citenamefont {Lewenstein}}]{Rana16}%
  \BibitemOpen
  \bibfield  {author} {\bibinfo {author} {\bibfnamefont {S.}~\bibnamefont
  {Rana}}, \bibinfo {author} {\bibfnamefont {P.}~\bibnamefont {Parashar}}, \
  and\ \bibinfo {author} {\bibfnamefont {M.}~\bibnamefont {Lewenstein}},\
  }\bibfield  {title} {\enquote {\bibinfo {title} {Trace-distance measure of
  coherence},}\ }\href {http://link.aps.org/doi/10.1103/PhysRevA.93.012110}
  {\bibfield  {journal} {\bibinfo  {journal} {Phys. Rev. A}\ }\textbf {\bibinfo
  {volume} {93}},\ \bibinfo {pages} {012110} (\bibinfo {year}
  {2016})}\BibitemShut {NoStop}%
\bibitem [{\citenamefont {Brito}\ \emph {et~al.}(2017)\citenamefont {Brito},
  \citenamefont {Amaral},\ and\ \citenamefont {Chaves}}]{Brito17}%
  \BibitemOpen
  \bibfield  {author} {\bibinfo {author} {\bibfnamefont {S.~G.~A.}\
  \bibnamefont {Brito}}, \bibinfo {author} {\bibfnamefont {B.}~\bibnamefont
  {Amaral}}, \ and\ \bibinfo {author} {\bibfnamefont {R.}~\bibnamefont
  {Chaves}},\ }\bibfield  {title} {\enquote {\bibinfo {title} {Quantifying bell
  non-locality with the trace distance},}\ }\href {https://link.aps.org/doi/10.1103/PhysRevA.97.022111}
  {\bibfield  {journal} {\bibinfo  {journal} {Phys. Rev. A}\ }\textbf {\bibinfo
  {volume} {97}},\ \bibinfo {pages} {022111} (\bibinfo {year}
  {2018})}\BibitemShut {NoStop}%
\bibitem [{\citenamefont {Vandenberghe}\ and\ \citenamefont
  {Boyd}(1996)}]{SDP}%
  \BibitemOpen
  \bibfield  {author} {\bibinfo {author} {\bibfnamefont {L.}~\bibnamefont
  {Vandenberghe}}\ and\ \bibinfo {author} {\bibfnamefont {S.}~\bibnamefont
  {Boyd}},\ }\bibfield  {title} {\enquote {\bibinfo {title} {Semidefinite
  programming},}\ }\href {http://dx.doi.org/10.1137/1038003} {\bibfield
  {journal} {\bibinfo  {journal} {SIAM Review}\ }\textbf {\bibinfo {volume}
  {38}},\ \bibinfo {pages} {49--95} (\bibinfo {year} {1996})}\BibitemShut
  {NoStop}%
\bibitem [{\citenamefont {Bavaresco}\ \emph {et~al.}(2017)\citenamefont
  {Bavaresco}, \citenamefont {Quintino}, \citenamefont {Guerini}, \citenamefont
  {Maciel}, \citenamefont {Cavalcanti},\ and\ \citenamefont
  {Cunha}}]{Bavaresco17}%
  \BibitemOpen
  \bibfield  {author} {\bibinfo {author} {\bibfnamefont {J.}~\bibnamefont
  {Bavaresco}}, \bibinfo {author} {\bibfnamefont {M.~T.}\ \bibnamefont
  {Quintino}}, \bibinfo {author} {\bibfnamefont {L.}~\bibnamefont {Guerini}},
  \bibinfo {author} {\bibfnamefont {T.~O.}\ \bibnamefont {Maciel}}, \bibinfo
  {author} {\bibfnamefont {D.}~\bibnamefont {Cavalcanti}}, \ and\ \bibinfo
  {author} {\bibfnamefont {Marcelo~T.}\ \bibnamefont {Cunha}},\ }\bibfield
  {title} {\enquote {\bibinfo {title} {Most incompatible measurements for
  robust steering tests},}\ }\href {\doibase 10.1103/PhysRevA.96.022110}
  {\bibfield  {journal} {\bibinfo  {journal} {Phys. Rev. A}\ }\textbf {\bibinfo
  {volume} {96}},\ \bibinfo {pages} {022110} (\bibinfo {year}
  {2017})}\BibitemShut {NoStop}%
\bibitem [{\citenamefont {Werner}(1989)}]{Werner89}%
  \BibitemOpen
  \bibfield  {author} {\bibinfo {author} {\bibfnamefont {R.~F.}\ \bibnamefont
  {Werner}},\ }\bibfield  {title} {\enquote {\bibinfo {title} {Quantum states
  with {E}instein-{P}odolsky-{R}osen correlations admitting a hidden-variable
  model},}\ }\href {http://link.aps.org/doi/10.1103/PhysRevA.40.4277}
  {\bibfield  {journal} {\bibinfo  {journal} {Phys. Rev. A}\ }\textbf {\bibinfo
  {volume} {40}},\ \bibinfo {pages} {4277--4281} (\bibinfo {year}
  {1989})}\BibitemShut {NoStop}%
\bibitem [{\citenamefont {Miranowicz}\ \emph {et~al.}(2008)\citenamefont
  {Miranowicz}, \citenamefont {Ishizaka}, \citenamefont {Horst},\ and\
  \citenamefont {Grudka}}]{Adam08}%
  \BibitemOpen
  \bibfield  {author} {\bibinfo {author} {\bibfnamefont {A.}~\bibnamefont
  {Miranowicz}}, \bibinfo {author} {\bibfnamefont {S.}~\bibnamefont
  {Ishizaka}}, \bibinfo {author} {\bibfnamefont {B.}~\bibnamefont {Horst}}, \
  and\ \bibinfo {author} {\bibfnamefont {A.}~\bibnamefont {Grudka}},\
  }\bibfield  {title} {\enquote {\bibinfo {title} {Comparison of the relative
  entropy of entanglement and negativity},}\ }\href
  {http://link.aps.org/doi/10.1103/PhysRevA.78.052308} {\bibfield  {journal}
  {\bibinfo  {journal} {Phys. Rev. A}\ }\textbf {\bibinfo {volume} {78}},\
  \bibinfo {pages} {052308} (\bibinfo {year} {2008})}\BibitemShut {NoStop}%
\bibitem [{\citenamefont {Vedral}\ and\ \citenamefont
  {Plenio}(1998)}]{Vedral98}%
  \BibitemOpen
  \bibfield  {author} {\bibinfo {author} {\bibfnamefont {V.}~\bibnamefont
  {Vedral}}\ and\ \bibinfo {author} {\bibfnamefont {M.~B.}\ \bibnamefont
  {Plenio}},\ }\bibfield  {title} {\enquote {\bibinfo {title} {Entanglement
  measures and purification procedures},}\ }\href
  {http://dx.doi.org/10.1103/PhysRevA.57.1619} {\bibfield  {journal} {\bibinfo
  {journal} {Phys. Rev. A}\ }\textbf {\bibinfo {volume} {57}},\ \bibinfo
  {pages} {1619} (\bibinfo {year} {1998})}\BibitemShut {NoStop}%
\bibitem [{\citenamefont {Fuchs}\ and\ \citenamefont {van~de
  Graaf}(1999)}]{Fuchs99}%
  \BibitemOpen
  \bibfield  {author} {\bibinfo {author} {\bibfnamefont {C.~A.}\ \bibnamefont
  {Fuchs}}\ and\ \bibinfo {author} {\bibfnamefont {J.}~\bibnamefont {van~de
  Graaf}},\ }\bibfield  {title} {\enquote {\bibinfo {title} {Cryptographic
  distinguishability measures for quantum-mechanical states},}\ }\href
  {\doibase 10.1109/18.761271} {\bibfield  {journal} {\bibinfo  {journal} {IEEE
  Trans. Inf. Th.}\ }\textbf {\bibinfo {volume} {45}},\ \bibinfo {pages}
  {1216--1227} (\bibinfo {year} {1999})}\BibitemShut {NoStop}%
\bibitem [{Note1()}]{Note1}%
  \BibitemOpen
  \bibinfo {note} {We need to fix the number of measurements on Alice's side to
  be able to define some steerability properties.}\BibitemShut {Stop}%
\bibitem [{\citenamefont {Ruskai}(1994)}]{RUSKAI94}%
  \BibitemOpen
  \bibfield  {author} {\bibinfo {author} {\bibfnamefont {M}~\bibnamefont
  {Ruskai}},\ }\bibfield  {title} {\enquote {\bibinfo {title} {Beyond strong
  subadditivity? improved bounds on the contraction of generalized relative
  entropy},}\ }\href {\doibase 10.1142/S0129055X94000407} {\bibfield  {journal}
  {\bibinfo  {journal} {Rev. Math. Phys.}\ }\textbf {\bibinfo {volume} {06}},\
  \bibinfo {pages} {1147--1161} (\bibinfo {year} {1994})}\BibitemShut {NoStop}%
\end{thebibliography}

%

\end{document}